\PassOptionsToPackage{pdfpagelabels=false}{hyperref}
\documentclass[fleqn,usenatbib,usedcolumn]{mnras}
\usepackage[british]{babel}             %
\usepackage{txfonts}                  %
\usepackage{graphicx}	%
\usepackage{hyperref} %
\usepackage{natbib}
\usepackage{aastexmacros}
\usepackage{tikz}
\usepackage[algoruled]{algorithm2e}
\usepackage[caption=false]{subfig}
\usepackage[mediumspace,mediumqspace,squaren]{SIunits}
\usepackage{pdflscape}

\usetikzlibrary{shapes,arrows,calc,positioning}

\renewcommand{\vec}[1]{\mathbf{#1}}
\newcommand{\jansky}{\text{Jy}}

\makeatletter
\newcommand\footnoteref[1]{\protected@xdef\@thefnmark{\ref{#1}}\@footnotemark}
\makeatother

\newcommand{\aref}[1]{\hyperref[#1]{Appendix~\ref{#1}}}
\newcommand{\eref}[1]{\hyperref[#1]{Equation~\ref{#1}}}

\AtBeginDocument{}%
\AtBeginDocument{}%
\AtBeginDocument{}%
\AtBeginDocument{}%

\title[Machine learning for radio cross-identification]{Radio Galaxy Zoo: Machine learning for radio source host galaxy cross-identification}

\author[Alger et al.]{M.~J.~Alger$^{1, 2}$\thanks{Email: \href{mailto:matthew.alger@anu.edu.au}{matthew.alger@anu.edu.au}},
  J.~K.~Banfield$^{1, 3}$,
  C.~S.~Ong$^{2, 4}$,
  L.~Rudnick$^{5}$,
  O.~I.~Wong$^{6, 3}$,
  C.~Wolf$^{1, 3}$,
  \newauthor
  H.~Andernach$^{7}$,
  R.~P.~Norris$^{8, 9}$,
  S.~S.~Shabala$^{10}$
\\
$^{1}$Research School of Astronomy and Astrophysics, The Australian National University, Canberra, ACT 2611, Australia\\
$^{2}$Data61, CSIRO, Canberra, ACT 2601, Australia\\
$^{3}$ARC Centre of Excellence for All-Sky Astrophysics (CAASTRO)\\
$^{4}$Research School of Computer Science, The Australian National University, Canberra, ACT 2601, Australia\\
$^{5}$Minnesota Institute for Astrophysics, University of Minnesota, 116 Church St. SE, Minneapolis, MN 55455\\
$^{6}$International Centre for Radio Astronomy Research-M468, The University of Western Australia, 35 Stirling Hwy, Crawley, WA 6009, Australia\\
$^{7}$Departamento de Astronom\'ia, DCNE, Universidad de Guanajuato, Apdo. Postal 144, CP 36000, Guanajuato, Gto., Mexico\\
$^{8}$Western Sydney University, Locked Bag 1797, Penrith South, NSW 1797, Australia\\
$^{9}$CSIRO Astronomy \& Space Science, PO Box 76, Epping, NSW 1710, Australia\\
$^{10}$School of Natural Sciences, University of Tasmania, Private Bag 37, Hobart, Tasmania 7001, Australia
}

\date{Accepted XXX. Received XXX}

\pubyear{2017}

\begin{document}
\label{firstpage}
\pagerange{\pageref{firstpage}--\pageref{lastpage}}
\maketitle

\begin{abstract}
  We consider the problem of determining the host galaxies of radio sources by
  cross-identification. This has traditionally been done manually, which will be intractable for wide-area radio surveys
  like the Evolutionary Map of the Universe (EMU). Automated
  cross-identification will be critical for these future surveys, and machine
  learning may provide the tools to develop such methods. We apply a standard
  approach from computer vision to cross-identification, introducing one
  possible way of automating this problem, and explore the pros and cons of
  this approach. We apply our method to the $1.4$~GHz Australian Telescope
  Large Area Survey (ATLAS) observations of the \emph{Chandra} Deep Field
  South (CDFS) and the ESO Large Area ISO Survey South 1 (ELAIS-S1) fields by
  cross-identifying them with the \emph{Spitzer} Wide-area Infrared
  Extragalactic (SWIRE) survey. We train our method with two sets of
  data: expert cross-identifications of CDFS from the initial ATLAS data
  release and crowdsourced cross-identifications of CDFS from Radio Galaxy
  Zoo. We found that a simple strategy of cross-identifying a radio component
  with the nearest galaxy performs comparably to our more complex methods, though our estimated best-case performance is near 100 per cent.
  ATLAS contains 87 complex radio sources that have been cross-identified by experts,
  so there are not enough complex examples to learn how to cross-identify them accurately. Much larger
  datasets are therefore required for training methods like ours. We also show
  that training our method on Radio Galaxy Zoo cross-identifications gives
  comparable results to training on expert cross-identifications,
  demonstrating the value of crowdsourced training data.
\end{abstract}


\section{Introduction}\label{introduction}

    \begin{figure*}
        \centering
        \subfloat[Two compact components, each a compact source.]{\includegraphics[width=0.30\linewidth]{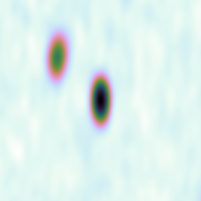}}\quad
        \subfloat[One resolved component and resolved source.]{\includegraphics[width=0.30\linewidth]{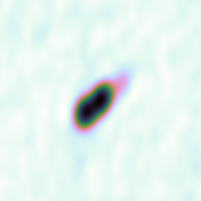}}\quad
        \subfloat[Three resolved components comprising one resolved source.]{\includegraphics[width=0.30\linewidth]{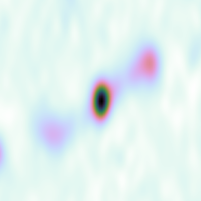}}
        \caption{\label{fig:definitions} Examples showing key definitions of radio emission regions used throughout this paper.
                 Compact and resolved components are defined by \autoref{eq:compact}.}
    \end{figure*}

  Next generation radio telescopes such as the Australian SKA Pathfinder
  \citep[ASKAP;][]{johnston07} and Apertif \citep{verheijen08} will conduct
  increasingly wide, deep, and high-resolution radio surveys, producing large
  amounts of data. The Evolutionary Map of the Universe
  \citep[EMU;][]{norris11} survey using ASKAP is expected to detect over 70 million
  radio sources, compared to the 2.5 million radio sources currently
  known \citep{banfield15}. An important part of processing these data is cross-identifying observed
  radio emission regions with observations of their host galaxy in surveys at
  other wavelengths.

  In the presence of extended radio emission cross-identification of the host can be a
  difficult task. Radio emission may extend far from the host galaxy
  and emission regions from a single physical object may appear disconnected. As a result, the
  observed structure of a radio source may have a complex relationship
  with the corresponding host galaxy, and cross-identification in radio is
  much more difficult than cross-identification at shorter wavelengths. Small surveys
  containing a few thousand sources such as the Australia Telescope Large Area Survey
  \citep[ATLAS;][]{norris06, middelberg08} can be cross-identified manually,
  but this is impractical for larger surveys.

  One approach to cross-identification of large numbers of sources is crowdsourcing, where volunteers
  cross-identify radio sources with their host galaxy. This is the premise of Radio Galaxy
  Zoo\footnote{\url{https://radio.galaxyzoo.org}} \citep{banfield15}, a
  citizen science project hosted on the Zooniverse platform \citep{lintott08}.
  Volunteers are shown radio and infrared images and are asked to
  cross-identify radio sources with the corresponding infrared host galaxies. An
  explanation of the project can be found in \citet{banfield15}. The first
  data release for Radio Galaxy Zoo will provide a large dataset of over
  75~000 radio-host cross-identifications and radio source morphologies
  (Wong et al., in prep). While this is a much larger number of visual
  cross-identifications than have been made by experts \citep[e.g.,
  ][]{Taylor2007,Gendre2008,Grant2010,norris06,middelberg08} it is still far
  short of the millions of radio sources expected to be detected in upcoming
  radio surveys \citep{norris17surveys}.

  Automated algorithms have been developed for cross-identification.
  \citet{fan15} applied Bayesian
  hypothesis testing to this problem, fitting a three-component model to extended radio
  sources. This was achieved under the assumption that extended radio sources
  are composed of a core radio component and two lobe components. The core
  radio component is coincident with the host galaxy, so cross-identification
  amounts to finding the galaxy coincident with the core radio component in
  the most likely model fit. This method is easily extended to use other, more
  complex models, but it is purely geometric. It does not incorporate
  other information such as the physical properties of the potential host
  galaxy. Additionally, there may be new classes of radio source detected in
  future surveys like EMU which do not fit the model. \citet{weston18lrpy}
  developed a modification of the likelihood ratio method of
  cross-identification \citep{richter75likelihood} for application to ATLAS
  and EMU. This method does well on non-extended radio sources
  with approximately 70 per cent accuracy in the ATLAS fields, but does
  not currently handle more complex (extended or multi-component) radio sources
  \citep{norris17unexpected}.

  One possibility is that machine learning techniques can
  be developed to automatically cross-identify catalogues drawn from new surveys. Machine learning
  describes a class of methods that learn approximations to functions. If
  cross-identification can be cast as a function approximation problem, then machine learning will allow data
  sets such as Radio Galaxy Zoo to be generalised to work on new data. Data sets from
  citizen scientists have already been used to train machine learning methods.
  Some astronomical examples can be found in \citet{marshall15citizenscience}.

  In this paper we cast cross-identification as a function
  approximation problem by applying an approach from computer vision
  literature. This approach casts cross-identification as the standard machine
  learning problem of binary classification by asking whether a given
  infrared source is the host galaxy or not. We train our methods on expert
  cross-identifications and volunteer cross-identifications from Radio Galaxy Zoo. In
  \autoref{sec:data} we describe the data we use to train our methods. In
  \autoref{sec:method} we discuss how we cast the radio host galaxy
  cross-identification problem as a machine learning problem. In
  \autoref{sec:results} we present results of applying our method to ATLAS
  observations of the \emph{Chandra} Deep Field South (CDFS) and the ESO Large Area ISO Survey South 1 (ELAIS-S1) field. Our data, code and results are
  available at \url{https://radiogalaxyzoo.github.io/atlas-xid}.

  Throughout this paper, a `radio source' refers to all radio emission observed associated with a single host galaxy, and a `radio component' refers to a single, contiguous
  region of radio emission. Multiple components may arise from a single
  source. A `compact' source is composed of a single unresolved component. \autoref{eq:compact} shows the definition of a resolved component. We
  assume that all unresolved components are compact sources, i.e. we assume that each unresolved component has its own host galaxy\footnote{This will be incorrect if the unresolved components are actually compact lobes or hotspots, but determining which components correspond to unique radio sources is outside the scope of this paper.}. An `extended'
  source is a non-compact source, i.e. resolved single-component sources or a
  multi-component source. \autoref{fig:definitions} illustrates these definitions.

\section{Data}\label{sec:data}

  We use radio data from the Australia Telescope Large Area Survey
  \citep[ATLAS;][]{norris06,franzen15}, infrared data from the \emph{Spitzer}
  Wide-area Infrared Extragalactic survey \citep[SWIRE;][]{lonsdale03swire,
  surace05swire}, and cross-identifications of these surveys from the citizen
  science project Radio Galaxy Zoo \citep{banfield15}. Radio Galaxy Zoo also
  includes cross-identifications of sources in Faint Images of the Radio Sky at
  Twenty-Centimeters \citep[FIRST;][]{white97first} and the All\emph{WISE}
  survey \citep{cutri2013wiseexplanatory}, though we focus only on Radio
  Galaxy Zoo data from ATLAS and SWIRE.

  \subsection{ATLAS}\label{sec:atlas}
    \begin{table}
      \caption{Catalogues of ATLAS/SWIRE cross-identifications for the CDFS
        and ELAIS-S1 fields. The method used to generate each catalogue is
        shown, along with the number of radio components cross-identified in each
        field.}
      \label{tab:atlas-cids}
      \begin{tabular}{llcc}
        \hline
        Catalogue & Method & CDFS & ELAIS-S1\\
        \hline
        \citet{norris06} & Manual & 784 & 0\\
        \citet{middelberg08} & Manual & 0 & 1366\\
        \citet{fan15} & Bayesian models & 784 & 0\\
        \citet{weston18lrpy} & Likelihood ratio & 3078 & 2113\\
        Wong et al. (in prep) & Crowdsourcing & 2460 & 0 \\
        \hline
      \end{tabular}
    \end{table}

    ATLAS is a pilot survey for the EMU \citep{norris11} survey, which will
    cover the entire sky south of $+30$ deg and is expected to detect
    approximately 70 million new radio sources. 95 per cent of these sources
    will be single-component sources, but the remaining 5 per cent pose a
    considerable challenge to current automated cross-identification methods
    \citep{norris11}. EMU will be conducted at the same depth and resolution
    as ATLAS, so methods developed for processing ATLAS data are expected to
    work for EMU. ATLAS is a wide-area radio survey of the CDFS and ELAIS-S1
    fields at 1.4~GHz with a sensitivity of 14 and
    \unit{17}{\micro\jansky}~beam$^{-1}$ on CDFS and ELAIS-S1 respectively.
    CDFS covers 3.6~deg$^2$ and contains 3034 radio components above a
    signal-to-noise ratio of 5. ELAIS-S1 covers 2.7~deg$^2$ and contains 2084
    radio components above a signal-to-noise ratio of 5 \citep{franzen15}. The
    images of CDFS and ELAIS-S1 have angular resolutions of 16 by 7 and 12 by
    8 arcsec respectively, with pixel sizes of 1.5 arcsec px$^{-1}$.
    \autoref{tab:atlas-cids} summarises catalogues that contain
    cross-identifications of radio components in ATLAS with host galaxies in
    SWIRE. In the present work, we train methods on
    CDFS\footnote{Radio Galaxy Zoo only contains CDFS sources and so
    we cannot train methods on ELAIS-S1.} and test these methods on both CDFS
    and ELAIS-S1. This ensures our methods are transferable to different
    areas of the sky observed by the same telescope as will be the case for
    EMU.

  \subsection{SWIRE}\label{sec:swire}

    SWIRE is a wide-area infrared
    survey at the four IRAC wavelengths 3.6, 4.5, 5.8, and
    \unit{8.0}{\micro\meter} \citep{lonsdale03swire, surace05swire}. It covers eight fields, including CDFS and ELAIS-S1. SWIRE is the source of infrared
    observations for cross-identification with ATLAS. SWIRE has catalogued 221,535
    infrared objects in CDFS and 186,059 infrared objects in ELAIS-S1 above a signal-to-noise ratio of 5.

  \subsection{Radio Galaxy Zoo}\label{sec:rgz}

    Radio Galaxy Zoo asks volunteers to cross-identify radio components with
    their infrared host galaxies. There are a total of 2460 radio components
    in Radio Galaxy Zoo sourced from ATLAS observations of CDFS. These components are
    cross-identified by Radio Galaxy Zoo participants with host galaxies
    detected in SWIRE. A more detailed description can be found in
    \citet{banfield15} and a full description of how the Radio Galaxy Zoo catalogue used in this work\footnote{The Radio Galaxy Zoo Data
    Release 1 catalogue will only include cross-identifications for which over
    65 per cent of volunteers agree. However, we use a preliminary catalogue containing volunteer
    cross-identifications for all components.} is generated can be found in Wong
    et al. (in prep).

    The ATLAS~CDFS radio components that appear in Radio Galaxy Zoo are drawn from a prerelease version of the third data release
    of ATLAS by \citet{franzen15}. In this release, each radio component was fit with a
    two-dimensional Gaussian. Depending on the residual of the fit, more than
    one Gaussian may be fit to one region of radio emission. Each of these
    Gaussian fits is listed as a radio component in the ATLAS component catalogue. The
    brightest radio component from the multiple-Gaussian fit is called the
    `primary component'. If there was only one Gaussian fit then this Gaussian is the primary component. Each primary component found in the ATLAS
    component catalogue appears in Radio Galaxy Zoo. Non-primary components
    may appear within the image of a primary component, but do not have their
    own entry in Radio Galaxy Zoo. We will henceforth only discuss the primary
    components.

  \section{Method}\label{sec:method}
    \begin{figure}
      \centering
      \includegraphics[width=\columnwidth]{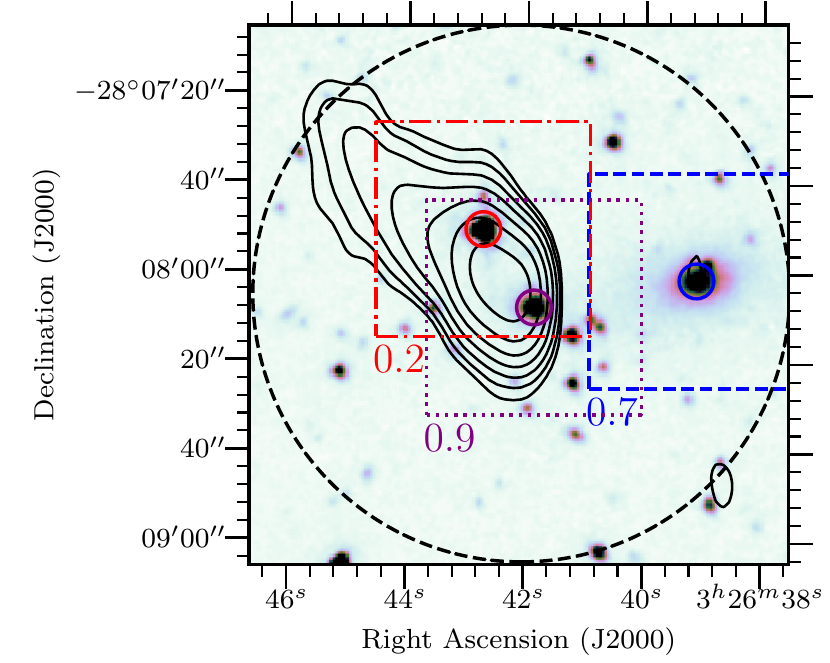}
      \caption{An example of finding the host galaxy of a radio source using
        our sliding-window method. The background image is a \unit{3.6}{\micro\meter} image from SWIRE. The contours show ATLAS radio data and start at $4\sigma$, increasing geometrically by a factor of 2. Boxes represent `windows'
        centred on candidate host galaxies, which are circled. The pixels in each window are used to represent the candidate that the window is centred
        on. The scores of each candidate would be calculated by a binary classifier using the window as input,
        and these scores are shown below each window. The scores
        shown are for illustration only. In this example,
        the galaxy coincident with the centre window would be chosen as the host galaxy, as this
        window has the highest score.
        The dashed circle
        shows the $1'$ radius from which candidate host galaxies are selected. For clarity, not all candidate host galaxies are shown.}
      \label{fig:windows}
    \end{figure}

  The aim of this paper is to express cross-identification in a form that
  will allow us to apply standard machine learning tools and methods. We use an approach from computer vision
  to cast cross-identification as binary classification.

  \subsection{Cross-identification as binary classification}\label{sec:xid-as-binary-classification}
    \begin{figure}
      \centering
      \tikzstyle{decision} = [diamond, draw, fill=white,
          text width=4.5em, text badly centered, inner sep=0pt]
      \tikzstyle{block} = [rectangle, draw, fill=white,
          text width=5em, text centered, rounded corners, minimum height=4em]
      \tikzstyle{line} = [draw, -latex']
      \begin{tikzpicture}[node distance=6mm, auto]
        \node [block] (init) {input radio source};
        \node [decision, right= of init] (iscompact) {compact?};
        \node [block, below= of iscompact] (compact) {find nearest infrared object};
        \node [block, right= of iscompact] (resolved) {find nearby infrared objects};
        \node [block, fill=black!30, right= of resolved] (classify) {classify objects};
        \node [block, below= of classify] (best) {choose object based on score};
        \coordinate (middle) at ($(compact)!0.5!(best)$);
        \node [block, below= of middle, fill=green!40] (done) {\textbf{host galaxy}};
        \path [line] (init) -- (iscompact);
        \path [line] (iscompact) -- (compact) node [midway] {yes};
        \path [line] (compact) -- (done);
        \path [line] (iscompact) -- (resolved) node [midway] {no};
        \path [line] (resolved) -- (classify);
        \path [line] (classify) -- (best);
        \path [line] (best) -- (done);
      \end{tikzpicture}
      \caption{Our cross-identification method once a binary classifier has been trained. As
        input we accept a radio component. If the component is compact, we assume it is a compact source and select
        the nearest infrared object as the host galaxy. If the component is
        resolved, we use the binary classifier to score all nearby infrared objects
        and select the highest-scored object as the host galaxy. Compact and resolved components are defined in \autoref{eq:compact}.}
      \label{fig:flowchart}
    \end{figure}
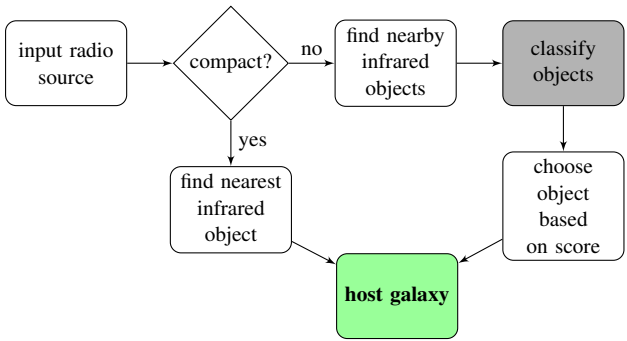

    We propose a two-step method for host galaxy cross-identification
    which we will describe now. Given a radio component, we want to find
    the corresponding host galaxy. The input is a $2' \times
    2'$ radio image of
    the sky centred on a radio component and potentially other information about
    objects in the image (such as the redshift or infrared colour). Images at other wavelengths (notably infrared) might be
    useful, but we defer this for now as it complicates the task.
    We chose a $2' \times 2'$ image to match the size of the images used
    by Radio Galaxy Zoo. To avoid solving the separate task of identifying
    which radio components are associated with the same source, we assume
    that each radio image represents a single extended
    source\footnote{Limitations of this assumption are discussed in
    \autoref{sec:limitations}.}. Radio cross-identification can then
    be formalised as follows: given a radio image centred on a radio
    component, locate the host galaxy of the source containing this radio
    component. This is a standard computer vision problem called `object
    detection', and we apply a common technique called a `sliding-window'
    \citep{rowley1996facedetection}.

    In sliding-window object detection, we want to find an object in an image.
    We develop a function to score each location in the image such that the
    highest-scored location coincides with the desired object (\eref{eq:score}). Square image
    cutouts called `windows' are taken centred on each location and these
    windows are used to represent that location in our scoring function. To
    find the infrared host galaxy, we choose the location with the highest
    score. To improve the efficiency of this process when applied to
    cross-identification, we only consider windows coincident with infrared
    sources detected in SWIRE. We call these infrared sources `candidate
    host galaxies'. For this paper, there is no use in scoring
    locations without infrared sources as that would not lead to a host identification
    anyway. Using candidate host galaxies instead of pixels also
    allows us to include ancillary information about the candidate host
    galaxies, such as their infrared colours and redshifts. We refer to the
    maximum distance a candidate host galaxy can be separated from a radio component as
    the `search radius' and take this radius to be $1$ arcmin. To score each
    candidate host galaxy we use a `binary classifier', which we will define
    now.

    \begin{algorithm}
        \KwData{\\\quad{}A $2 \times 2$ arcmin radio image of a radio component%
                \\\quad{}A set of infrared candidate host galaxies $\mathcal G$%
                \\\quad{}A binary classifier $f : \mathbb R^k \to \mathbb{R}$}
        \KwResult{A galaxy $g \in \mathcal G$}

        $max \leftarrow -\infty$\;
        $host \leftarrow \emptyset$\;
        \For{$g \in \mathcal G$}{
          $x \leftarrow$ a $k$-dimensional vector representation of $g$ (\autoref{vector-representation-of-infrared-sources})\;
          $d \leftarrow$ distance between $g$ and the radio component\;
          $score \leftarrow f(x) \times \frac{1}{\sqrt{2\pi\sigma^2}} \exp\left(-\frac{d^2}{2\sigma^2}\right);\quad \stepcounter{equation}\hfill(\theequation)\label{eq:score}$
          \BlankLine
          \If{$score > max$}{
            $max \leftarrow score$\;
            $host \leftarrow g$\;
          }
        }

        \KwRet{$host$}
        \caption{Cross-identifying a radio component given a radio image of the component, a catalogue of infrared candidate host galaxies and a binary classifier.
          $\sigma$ is a parameter of the method.}
        \label{alg:xid}
    \end{algorithm}

    Binary classification is a common method in machine learning
    where objects are to be assigned to one of two classes,
    called the `positive' and `negative' classes. This assignment is
    represented by the probability that an object is in the positive class. A
    `binary classifier' is a function mapping from an object to such a
    probability. Our formulation of cross-identification is equivalent to
    binary classification of candidate host galaxies: the positive class
    represents host galaxies, the negative class represents non-host galaxies,
    and to cross-identify a radio component we find the candidate host galaxy
    maximising the positive class probability. In other words,
    the binary classifier is exactly the sliding-window scoring function. We therefore split
    cross-identification into two separate tasks: the `candidate
    classification task' where, given a candidate host galaxy, we wish to
    determine whether it is a host galaxy of \emph{any} radio component; and
    the `cross-identification task' where, given a specific radio
    component, we wish to find its host galaxy. The candidate classification task
    is a traditional machine learning problem which results in a binary
    classifier. To avoid ambiguity and recognise that the values output by a
    binary classifier are not true probabilities, we will refer to the outputs
    of the binary classifier as `scores' in line with the sliding-window approach
    described above. The cross-identification task maximises over scores
    output by this classifier. Our approach is illustrated in
    \autoref{fig:windows} and described in \autoref{alg:xid}. We refer to the
    binary classifier scoring a candidate host galaxy as
    $f$. To implement $f$ as a function that accepts candidate host galaxies
    as input, we need to represent candidate host galaxies by vectors. We
    describe this in \autoref{vector-representation-of-infrared-sources}.
    There are many options for modelling $f$. In this paper we apply three
    different models: logistic regression, random forests and convolutional
    neural networks.

\begin{figure}
      \centering
      \includegraphics[width=\linewidth]{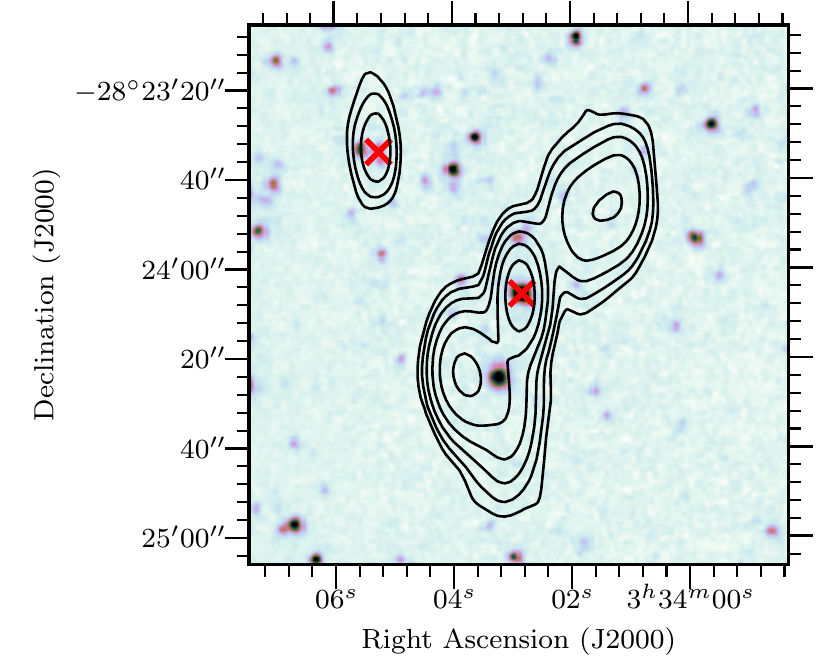}
      \caption{A $2'$-wide radio image centred on ATLAS3\textunderscore{}J033402.87-282405.8C.
        This radio source breaks the assumption that there are no other radio
        sources within 1~arcmin of the source. Another radio source is visible
        to the upper-left. Host galaxies found by Radio Galaxy Zoo volunteers
        are shown by crosses. The background image
        is a \unit{3.6}{\micro\meter} image from SWIRE. The contours show ATLAS radio data and start at $4\sigma$, increasing geometrically by a factor of 2.}
      \label{fig:broken-isolation}
    \end{figure}

    We cross-identify each radio component in turn. The classifier $f$
    provides a score for each candidate host galaxy. This score indicates how
    much the candidate looks like a host galaxy, independent of which radio
    component we are currently cross-identifying. If there are other nearby host
    galaxies, then multiple candidate hosts may have high scores (e.g.
    \autoref{fig:broken-isolation}). This difficulty is necessary --- a classifier
    with dependence on radio object would be impossible to train. We
    need multiple positive examples (i.e. host galaxies) to train a binary classifier, but
    for any specific radio component there is only one host galaxy. As a
    result, the candidate classification task aims to answer the general question
    of whether a given galaxy is the host galaxy of \emph{any} radio
    component, while the cross-identification task attempts to cross-identify
    a \emph{specific} radio component. To distinguish between candidate host
    galaxies with high scores, we weight the scores by a Gaussian function of
    angular separation between the candidates and the radio component. The
    width of the Gaussian, $\sigma$, controls the influence of the Gaussian on
    the final cross-identification. When $\sigma$ is small, our approach is
    equivalent to a nearest-neighbours approach where we select the nearest
    infrared object to the radio component as the host galaxy. In the limit
    where $\sigma \to \infty$, we maximise the score output by the
    classifier as above. We take $\sigma = 30''$ as this was the best value
    found by a grid search. Note that the optimum width will depend on
    the density of radio sources on the sky, the angular separation of the
    host galaxy and its radio components and the angular resolution of the survey.

    We can improve upon this method by cross-identifying compact radio sources
    separately from extended sources, as compact sources are much easier to
    cross-identify. For a compact source, the nearest SWIRE object may be
    identified as the host galaxy (a \emph{nearest-neighbours} approach), or a
    more complex method such as likelihood ratios may be applied
    \citep[see][]{weston18lrpy}. We cross-identify compact sources separately
    in our pipeline and this process is shown in \autoref{fig:flowchart}.

  \subsection{Limitations of our approach}
    \label{sec:limitations}

    \begin{figure}
      \centering
      \includegraphics[width=\linewidth]{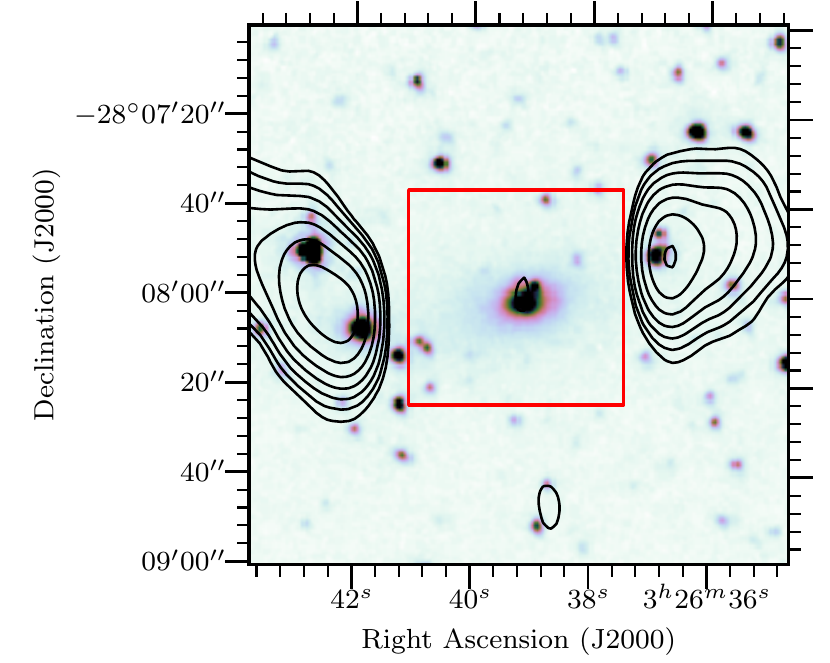}
      \caption{An example of a radio source where the window centred on the
        host galaxy, shown as a rectangle, does not contain enough radio
        information to correctly identify the galaxy as the host. The background image
        is a \unit{3.6}{\micro\meter} image from SWIRE. The contours show ATLAS radio data and start at $4\sigma$, increasing geometrically by a factor of 2.}
      \label{fig:broken-window-size}
    \end{figure}

    \begin{figure}
      \centering
      \includegraphics[width=\linewidth]{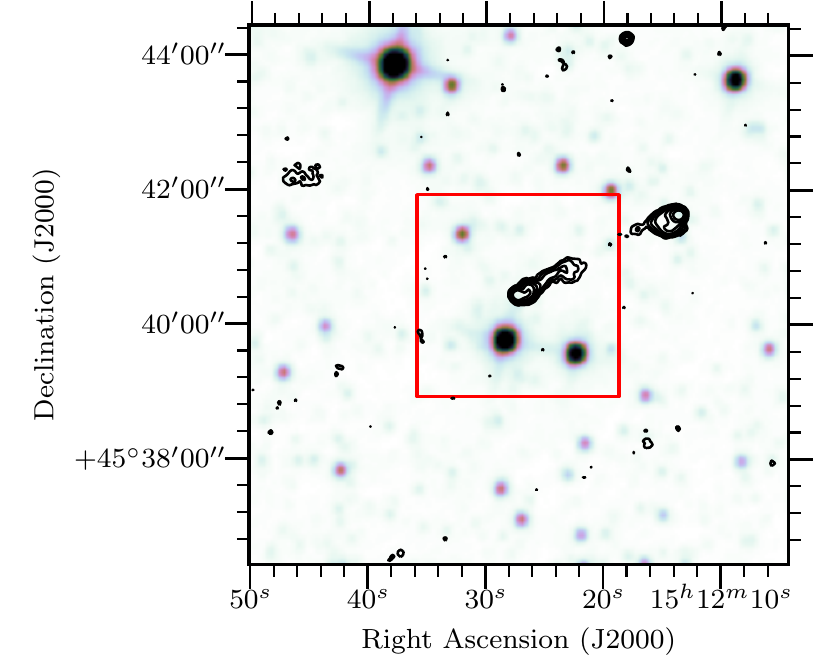}
      \caption{A $8'$-wide radio image from FIRST, centred on
        FIRST\ J151227.2+454026. The $3'$-wide red box indicates the boundaries of
        the image of this radio component shown to volunteers in Radio Galaxy
        Zoo. This radio source breaks our assumption that the whole radio source
        is visible in the chosen radius. As one of the components of the radio source
        is outside of the image, a volunteer (or automated algorithm) looking at
        the $3'$-wide image may be unable to determine that this is a radio
        double or locate the host galaxy. The background image
        is a \unit{3.4}{\micro\meter} image from \emph{WISE}. The contours show FIRST radio
        data, starting at $4\sigma$ and increasing geometrically by a factor of 2.}
      \label{fig:broken-contains}
    \end{figure}

    We make a number of assumptions to relate the cross-identification task to
    the candidate classification task:
    \begin{enumerate}
      \item For any radio component, the $2' \times 2'$ image centred on the component contains components of only one radio source.
      \item For any radio component, the $2' \times 2'$ image centred on the component contains all components of this source.
      \item The host galaxy of a radio component is within the 1~arcmin search radius around the component, measured from the centre of the Gaussian fit.
      \item The host galaxy of a radio component is closer on the sky to the
        radio component than the host galaxy of any other radio component.
      \item The host galaxy appears in the SWIRE catalogue.
    \end{enumerate}
    These assumptions limit the effectiveness of our approach, regardless of
    how accurate our binary classifier may be. Examples of radio sources that break these respective assumptions are:
    \begin{enumerate}
      \item A radio source less than $1'$ away from another radio source.
      \item A radio source with an angular size greater than $2'$.
      \item A radio source with a component greater than $1'$ away from the host galaxy.
      \item A two-component radio source with another host galaxy between a component and the true host galaxy.
      \item An infrared-faint radio source \citep[as in][]{collier14irfaint}.
    \end{enumerate}

    The main limitations are problems of scale in choosing the
    candidate search radius and the size of the windows
    representing candidates. If the search radius is too small, we may not
    consider the host galaxy as a candidate. If the search radius is too
    large, we may consider multiple host galaxies (though this is mostly
    mitigated by the Gaussian weighting). If the window is too small, radio
    emission may extend past the edges of the window and we may miss critical
    information required to identify the galaxy as a host galaxy. If the
    window is too large, then irrelevant information will be included and it
    may be difficult or computationally expensive to score. We chose a
    window size of $32 \times 32$ pixels, corresponding to approximately $48'' \times 48''$ in
    ATLAS. This is shown as squares in \autoref{fig:windows} and
    \autoref{fig:broken-window-size}. These kinds of size problems are
    difficult even for non-automated methods as radio sources can be extremely
    wide --- for example, Radio Galaxy Zoo found a radio giant that spanned
    over three different images presented to volunteers and the full source
    was only cross-identified by the efforts of citizen scientists
    \citep{banfield15}. An example of a radio image where part of the radio
    source is outside the search radius is shown in
    \autoref{fig:broken-contains}.

    In weighting the scores by a Gaussian function of angular
    separation, we implicitly assume that the host galaxy of a radio component
    is closer to that radio component than any other host galaxy. If this
    assumption is not true then the incorrect host galaxy may be identified, though
    this is rare.

    We only need to require that the host galaxy appears in SWIRE to
    incorporate galaxy-specific features
    (\autoref{vector-representation-of-infrared-sources}) and to improve
    efficiency. Our method is applicable even when host galaxies are not detected in
    the infrared by considering every pixel of the radio image as a candidate
    location as would be done in the original computer vision approach. If the host galaxy location does not correspond to an infrared source, the radio source would be classified as infrared-faint.

    Our assumptions impose an upper bound on how well we can cross-identify
    radio sources. We estimate this upper bound in \autoref{sec:cdfs-results}.

  \subsection{Feature vector representation of infrared sources}
  \label{vector-representation-of-infrared-sources}

    Inputs to binary classifiers must be represented by an array of real values called feature vectors. We therefore need to choose a feature vector representation of our candidate host galaxies. Candidate hosts are sourced from the SWIRE catalogue (\autoref{sec:swire}). We represent each candidate host with 1034 real-valued features, combining the windows centred on each candidate (\autoref{sec:xid-as-binary-classification}) with ancillary infrared data from the SWIRE catalogue. For a given candidate host, these features are:
    \begin{itemize}
      \item the 6 base-10 logarithms of the ratios of fluxes of the candidate
        host at the four IRAC wavelengths (the `colours' of the candidate);
      \item the flux of the host at \unit{3.6}{\micro\meter};
      \item the stellarity index of the host at both 3.6 and
        \unit{4.5}{\micro\meter};
      \item the radial distance between the candidate host and the nearest
        radio component in the ATLAS catalogue; and
      \item a 32 $\times$ 32 pixel image from ATLAS (approximately $48''
        \times 48''$), centred on the candidate host (the window).
    \end{itemize}

    The infrared colours provide insight into the properties of the candidate
    host galaxy \citep{grant11polarised}. The 3.6 and \unit{4.5}{\micro\meter} fluxes trace
    both galaxies with faint polycyclic aromatic hydrocarbon (PAH) emission (i.e. late-type, usually star-forming galaxies)
    and elliptical galaxies dominated by old stellar populations. The
    \unit{5.8}{\micro\meter} flux selects galaxies where the infrared emission
    is dominated by non-equilibrium emission of dust grains due to active galactic nuclei,
    while the \unit{8.0}{\micro\meter} flux
    traces strong PAH emission at low redshift \citep{Sajina2005}. The
    stellarity index is a value in the SWIRE catalogue that represents how likely the object is to be a star rather
    than a galaxy \citep{surace05swire}. It was estimated by a neural network in
    \texttt{SExtractor} \citep{bertin96sextractor}.

    We use the $32 \times 32$ pixels of each radio window as independent
    features for all binary classification models, with the convolutional neural
    network automatically extracting features that are relevant. Other
    features of the radio components may be used instead of just relying on the pixel values,
    but there has been limited research on extracting such features:
    \citet{proctor06} describes hand-selected features for radio doubles in
    FIRST, and \citet{aniyan17cnn} and \citet{lukic18compact} make use of
    deep convolutional neural networks which automatically extract features as
    part of classification. A more comprehensive investigation of features is
    a good avenue for potential improvement in our pipeline but this is beyond
    the scope of this initial study.

  \subsection{Binary Classifiers}\label{sec:classifiers}

    We use three different binary classification models: logistic regression,
    convolutional neural networks, and random forests. These models cover
    three different approaches to machine learning. Logistic regression is a
    probabilistic binary classification model. It is linear in the feature
    space and outputs the probability that the input has a positive
    label~\citep[Chap. 4]{bishop06ml}. Convolutional neural networks (CNN) are
    biologically-inspired prediction models with image inputs.
    They have recently produced good results on large image-based datasets in
    astronomy \citep[e.g.]{lukic18compact, dieleman15cnn}. Random
    forests are an ensemble of decision trees~\citep{breiman01random-forest}.
    They consider multiple subsamples of the training set, where each
    bootstrap subsample is sampled with replacement from the training set. To
    classify a new data point, the random forest takes the weighted average of
    all classifications produced by each decision tree.

    Further details and background of these models are presented in \aref{app:models}.

  \subsection{Labels}\label{sec:labels}

    \begin{figure}
      \centering
      \includegraphics[width=\columnwidth]{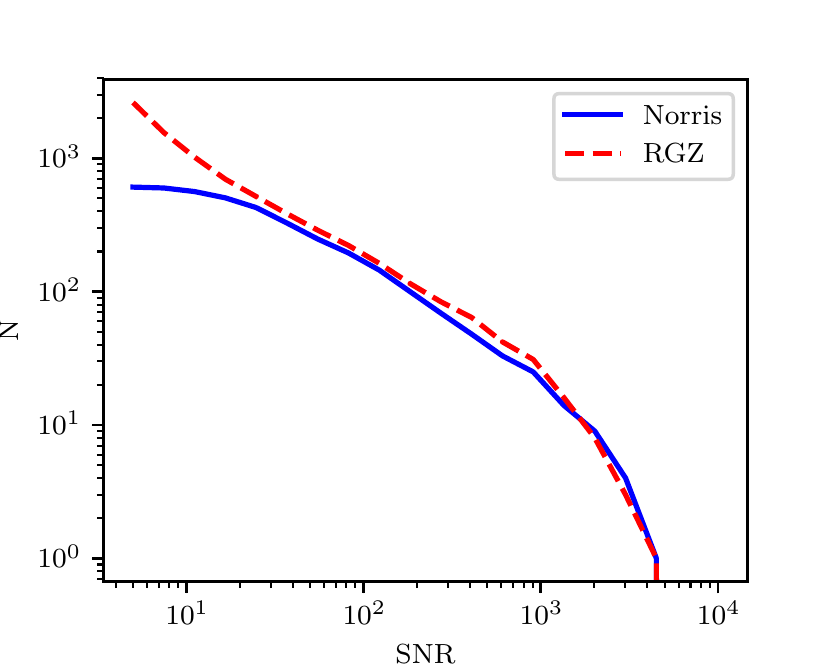}
      \caption{Cumulative number of radio components ($N$) in the expert (Norris) and Radio Galaxy
        Zoo (RGZ) training sets with different signal-to-noise ratios (SNR).}
      \label{fig:distribution-cutoffs}
    \end{figure}

    \begin{figure}
      \centering
      \includegraphics[width=\columnwidth, trim={0cm 0.5cm 0cm 0.5cm}, clip]{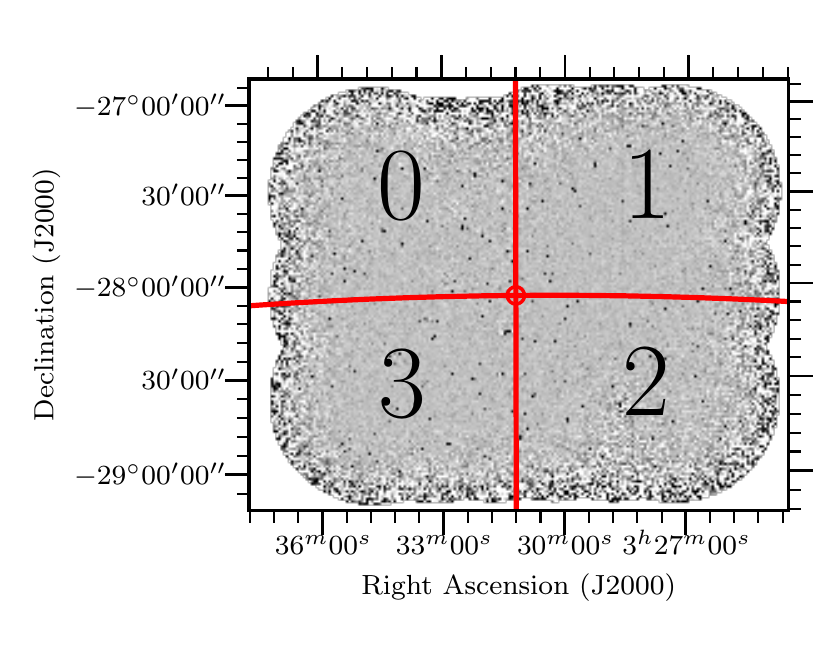}
      \caption{CDFS field training and testing quadrants labelled 0 -- 3. The
        central dot is located at $\alpha = 03^\text{h}31^\text{m}12^\text{s},
        \delta = -28^\circ{}06'00''$. The quadrants were chosen such that
        there are similar numbers of radio sources in each
        quadrant.\label{fig:quadrants}}
    \end{figure}

    \begin{table}
      \caption{Number of compact and resolved radio objects in each CDFS
      quadrant. Radio Galaxy Zoo (RGZ) has more cross-identifications than the
      expert catalogue \citep{norris06} provides as it uses a deeper data release of ATLAS, and
      so has more objects in each quadrant for training.}
      \label{tab:radio-count}
      \begin{tabular}{lcccc}
        \hline
        Quadrant & Compact & Resolved & Compact & Resolved\\
        &&&(RGZ)&(RGZ)\\
        \hline
        0 & 126 & 24 & 410 & 43 \\
        1 & 99 & 21 & 659 & 54 \\
        2 & 61 & 24 & 555 & 57 \\
        3 & 95 & 18 & 631 & 51 \\
        \hline
        \textit{Total} & 381 & 87 & 2255 & 205\\
        \hline
      \end{tabular}
    \end{table}

    The Radio Galaxy Zoo and \citet{norris06} cross-identification
    catalogues must be converted to binary labels for infrared objects so that
    they can be used to train binary classifiers. There are two challenges with this conversion:
    \begin{itemize}
      \item We can only say that an object is \emph{a} host galaxy, not which radio object it is associated with, and
      \item We cannot disambiguate between non-host infrared objects and host galaxies that were not in the cross-identification catalogue.
    \end{itemize}
    
    We use the Gaussian weighting described in \autoref{sec:xid-as-binary-classification} to address the first issue.
    The second issue is known as a `positive-unlabelled' classification problem, which is
    a binary classification problem where we only observe labels for the
    positive class. We treat unlabelled objects as negative examples following
    \citet{menon15cpe}. That is, we make the na\"ive assumption that any
    infrared object in the SWIRE catalogue not identified as a host galaxy in a cross-identification catalogue is not a host galaxy at all.

    We first generate positive labels from a cross-identification catalogue.
    We decide that if an infrared object is listed in the catalogue, then it
    is assigned a positive label as a host galaxy. We then assign every other galaxy a negative label. This has some problems
    --- an example is that if the cross-identification catalogue did not include a radio
    object (e.g.~it was below the signal-to-noise ratio) then the host galaxy
    of that radio object would receive a negative label. This occurs with
    \citet{norris06} cross-identifications, as these are associated with the
    first data release of ATLAS. The first data release went to a 5$\sigma$
    flux density level of $S_{1.4} \geq \unit{200}{\micro\jansky}\text{
    beam}^{-1}$ \citep{norris06}, compared to $S_{1.4} \geq \unit{85}{\micro\jansky}\text{
    beam}^{-1}$ for the third data release used by Radio Galaxy Zoo
    \citep{franzen15}. The labels from \citet{norris06} may therefore disagree with labels
    from Radio Galaxy Zoo even if they are both plausible. The difference in
    training set size at different flux cutoffs is shown in
    \autoref{fig:distribution-cutoffs}. We train and test our binary
    classifiers on infrared objects within a 1~arcmin radius of an ATLAS radio
    component.

  \subsection{Experimental Setup}
  \label{sec:experimental-setup}

    We trained binary classifiers on infrared objects in the CDFS field using
    two sets of labels. One label set was derived from Radio Galaxy Zoo
    cross-identifications and the other was derived from the \citet{norris06}
    cross-identification catalogue. We refer to these as the `Radio Galaxy Zoo
    labels' and the `expert labels' respectively. We divided the CDFS field
    into four quadrants for training and testing. The quadrants were divided
    with a common corner at $\alpha = 03^\text{h}31^\text{m}12^\text{s},
    \delta = -28^\circ{}06'00''$ as shown in \autoref{fig:quadrants}. For
    each trial, one quadrant was used to extract test examples and the other
    three quadrants were used for training examples.

    We further divided the radio components into compact and resolved. Compact
    components are cross-identified by fitting a 2D Gaussian \citep[as
    in][]{norris06} and we would expect any machine learning approach for host
    cross-identification to attain high accuracy on this set. A radio component was
    considered resolved if
    \begin{equation}
      \label{eq:compact}
        \ln \left(
          \frac{S_{\text{int}}}
               {S_{\text{peak}}}
        \right) > 2\sqrt{\left(
          \frac{\sigma_{S_{\text{int}}}}
               {S_{\text{int}}}
        \right)^2 + \left(
          \frac{\sigma_{S_{\text{peak}}}}
               {S_{\text{peak}}}
        \right)^2}\,\,\,\,,
    \end{equation}%
    where \(S_{\text{int}}\) is the integrated flux density,
    \(S_{\text{peak}}\) is the peak flux density, $\sigma_{S_{\text{int}}}$ is
    the uncertainty in integrated flux density and $\sigma_{S_{\text{peak}}}$
    is the uncertainty in peak flux density \citep[following][]{franzen15}.

    Candidate hosts were selected from the SWIRE catalogue. For a given subset
    of radio components, all SWIRE objects within 1~arcmin of all radio
    components in the subset were added to the associated SWIRE subset. In results
    for the candidate classification task, we refer to SWIRE objects
    within 1~arcmin of a compact radio component as part of the `compact set',
    and SWIRE objects within 1~arcmin of a resolved radio component as part of
    the `resolved set'.

    To reduce bias in the testing data due to the expert labels being
    generated from a shallower data release of ATLAS, a SWIRE object was only
    included in the test set if it was within 1~arcmin of a radio object with
    a SWIRE cross-identification in both the \citet{norris06} catalogue and the
    Radio Galaxy Zoo catalogue.

    Each binary classifier was trained on the training examples and used to
    score the test examples. These scores were thresholded to generate labels which could be directly compared
    to the expert labels. We then computed the `balanced accuracy' of these predicted labels. Balanced
    accuracy is the average of the accuracy on the positive class and the
    accuracy on the negative class, and is not sensitive to class imbalance.
    The candidate classification task has highly imbalanced classes --- in our
    total set of SWIRE objects within 1~arcmin of an ATLAS object, only 4 per
    cent have positive labels. Our threshold was chosen to maximise the balanced
    accuracy on predicted labels of the training set. Only examples within 1~arcmin of ATLAS objects
    in the first ATLAS data release \citep{norris06} were used to compute
    balanced accuracy, as these were the only ATLAS objects with expert labels.

    We then used the scores to predict the host galaxy
    for each radio component cross-identified by both \citet{norris06} and
    Radio Galaxy Zoo. We followed \autoref{alg:xid}:
    the score of each SWIRE object within 1~arcmin of a given radio
    component was weighted by a Gaussian function of angular separation from the
    radio component and the object with the highest
    weighted score was chosen as the host galaxy. The
    cross-identification accuracy was then estimated as the
    fraction of the predicted host galaxies that matched the \citet{norris06}
    cross-identifications.

\section{Results}\label{sec:results}

  In this section we present accuracies of our method trained on CDFS and
  applied to CDFS and ELAIS-S1, as well as results motivating our accuracy
  measures and estimates of upper and lower bounds for cross-identification
  accuracy using our method.

  \subsection{Application to ATLAS-CDFS}
  \label{sec:cdfs-results}

    \begin{figure}
      \centering
      \includegraphics[width=\columnwidth]{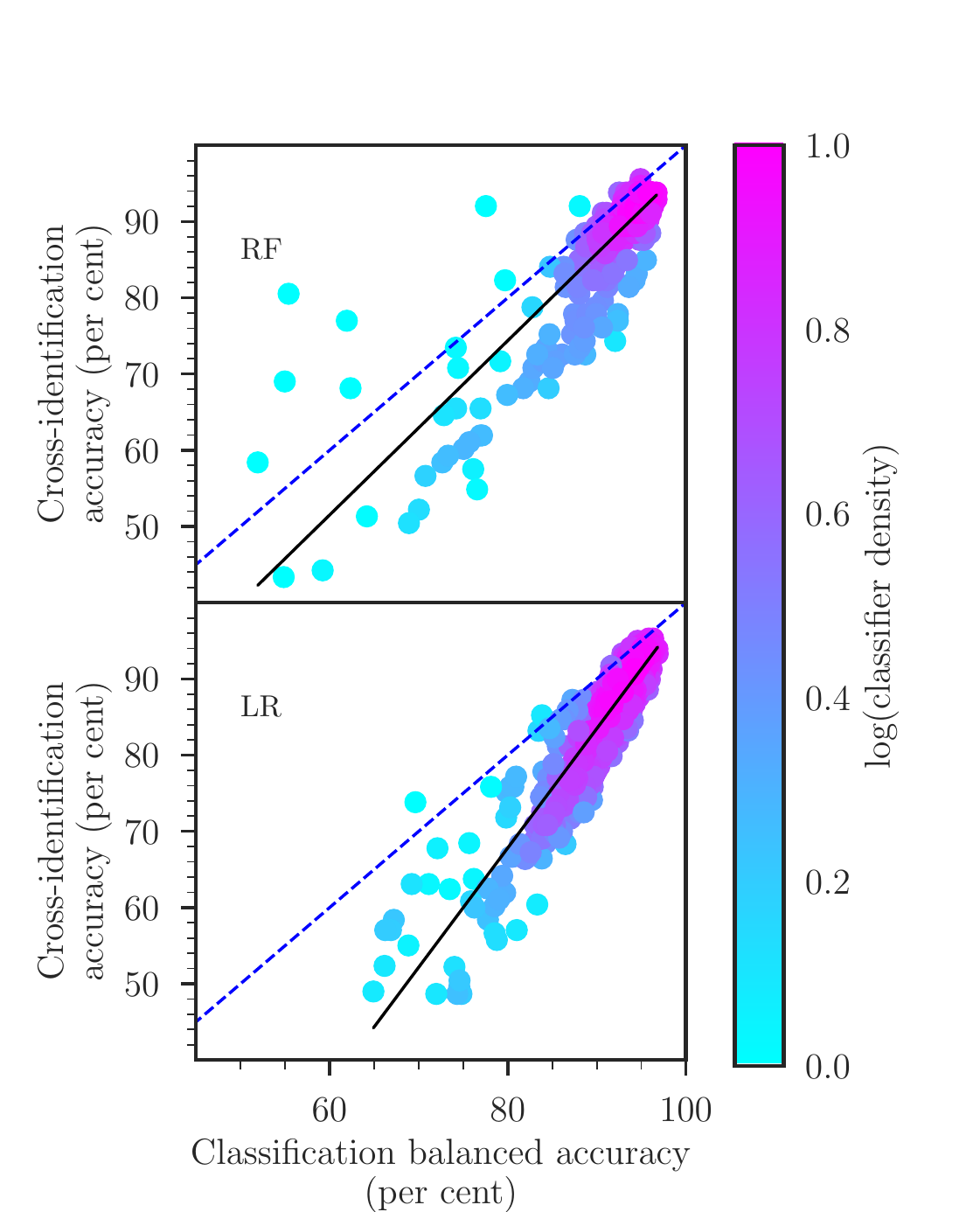}
      \caption{Balanced accuracy on the candidate classification task plotted
      against accuracy on the cross-identification task. `RF' indicates
      results from random forests, and `LR' indicates results from logistic
      regression. Binary classifiers were trained on random, small subsets of the
      training data to artificially restrict their accuracies. Colour shows
      the density of points on the plot estimated by a Gaussian kernel density
      estimate. The solid lines indicate the best linear fit; these fits have
      $R^2 = 0.92$ for logistic regression and $R^2 = 0.87$ for random
      forests.
      The dashed line shows the line where cross-identification accuracy and candidate classification accuracy are equal.
      We did not include convolutional neural networks in this test,
      as training them is very computationally expensive. There are 640 trials shown per classification model. These results
      exclude binary classifiers with balanced accuracies less than 51 per cent, as
      these are essentially random.
      \label{fig:gct-to-xid}}
    \end{figure}

    \begin{figure}
      \centering
      \includegraphics[width=\columnwidth]{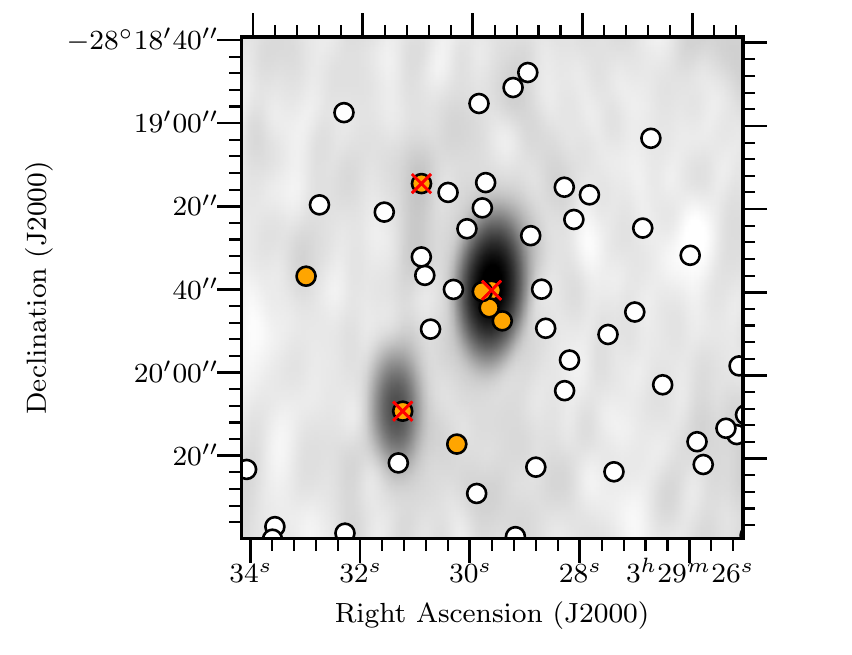}
      \caption{Predicted host galaxies in the candidate classification task for ATLAS3~J032929.61-281938.9. The background image is an ATLAS radio image. Radio Galaxy Zoo host galaxies
      are marked by crosses. SWIRE candidate host galaxies are circles coloured by the score output by a logistic regression binary classifier. The scores are thresholded to obtain labels, as when we compute balanced accuracy. Orange circles have been assigned a `positive' label by a logistic regression binary classifier and white otherwise. Note that there are more predicted host galaxies than there are radio components, so not all of the predicted host galaxies would be assigned as host galaxies in the cross-identification task.
      \label{fig:positives}}
    \end{figure}

    \begin{figure*}
    \centering
    \includegraphics[]{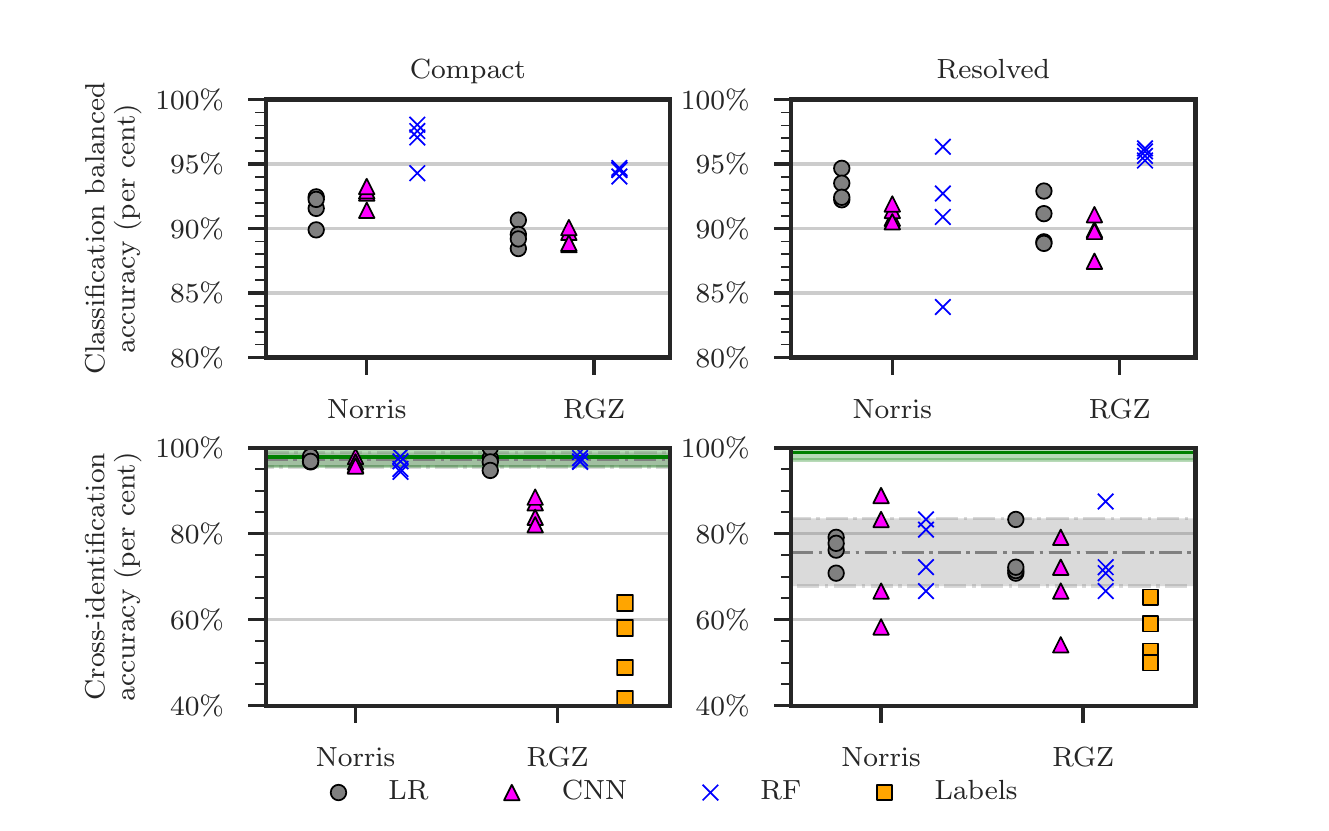}
    \caption{Performance of our method with logistic regression (`LR'), convolutional neural networks (`CNN') and random forest (`RF') binary classifiers. `Norris' indicates the performance of binary classifiers trained on the expert labels and `RGZ' indicates the performance of binary classifiers trained on the Radio Galaxy Zoo labels. One point is shown per binary classifier per testing quadrant. The training and testing sets have been split into compact (left) and resolved (right) objects. Shown for comparison is the accuracy of the Radio Galaxy Zoo consensus cross-identifications on the cross-identification task, shown as `Labels'. The cross-identification accuracy attained by a perfect binary classifier is shown by a solid green line, and the cross-identification accuracy of a nearest-neighbours approach is shown by a dashed grey line. The standard deviation of these accuracies across the four CDFS quadrants is shown by the shaded area. Note that the pipeline shown in \autoref{fig:flowchart} is not used for these results. \label{fig:ba}}
    \end{figure*}

    \begin{figure}
      \centering
      \includegraphics[]{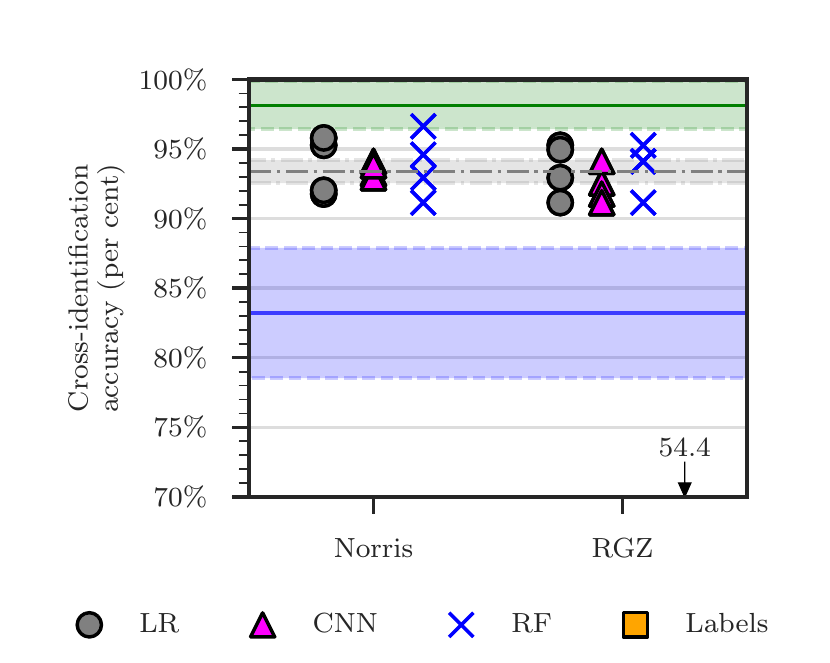}
      \caption{Performance of our approach using different binary classifiers on the cross-identification task. Markers and lines are as in \autoref{fig:ba}. The blue solid line indicates the performance of a random binary classifier and represents the minimum accuracy we expect to obtain. The standard deviation of this accuracy across 25 trials and 4 quadrants is shaded. The accuracy of Radio Galaxy Zoo on the cross-identification task is below the axis and is instead marked by an arrow with the mean accuracy. Note that the pipeline shown in \autoref{fig:flowchart} is used here, so compact objects are cross-identified in the same way regardless of binary classifier model. \label{fig:cross-id-accuracy}}
    \end{figure}

    We can assess trained binary classifiers either by their performance on
    the candidate classification task or by their performance on the
    cross-identification task when used in our method. Both performances are
    useful: performance on the candidate classification task provides a robust
    and simple way to compare binary classifiers without the limitations of
    our specific formulation, and performance on the cross-identification task
    can be compared with other cross-identification methods. We therefore
    report two sets of accuracies: balanced accuracy for the galaxy
    classification task and accuracy for the cross-identification task. These
    accuracy measures are correlated and we show this correlation in
    \autoref{fig:gct-to-xid}. Fitting a line of best fit with \texttt{scipy}
    gives $R^2 = 0.92$ for logistic regression and $R^2 = 0.87$ for random
    forests. While performance on the candidate classification task is correlated
    with performance on the cross-identification task, balanced accuracy does
    not completely capture the effectiveness of a binary classifier applied to
    the cross-identification task. This is because while our binary
    classifiers output real-valued scores, these scores are thresholded to
    compute the balanced accuracy. In the candidate classification
    task, the binary classifier only needs to ensure that host galaxies are
    scored higher than non-host galaxies. This means
    that after thresholding there can be
    many `false positives' that do not affect cross-identification. An example
    of this is shown in \autoref{fig:positives}, where the classifier has
    identified 8 `host galaxies'. However, there are only three true host
    galaxies in this image --- one per radio component --- and so in the
    cross-identification task, only three of these galaxies will be identified
    as hosts.

    In \autoref{fig:ba} we plot the balanced accuracies of our classification models
    on the candidate classification task and the cross-identification
    accuracies of our method using each of these models. Results are shown for both
    the resolved and compact sets. For comparison, we also plot the cross-identification accuracy of Radio Galaxy
    Zoo and a nearest-neighbours approach, as well as estimates for upper and
    lower limits on the cross-identification accuracy. We estimate the upper limit on performance by assigning all
    true host galaxies a score of 1 and
    assigning all other candidate host galaxies a score of 0. This
    is equivalent to `perfectly' solving the candidate classification task and so
    represents the best possible cross-identification performance achievable
    with our method. We estimate the lower limit on performance by 
    assigning random scores to each candidate host galaxy. We expect any
    useful binary classifier to produce better
    results than this, so this represents the lowest expected
    cross-identification performance. The upper estimates, lower estimates,
    and nearest-neighbour accuracy are shown as horizontal lines in
    \autoref{fig:ba}.

    In \autoref{fig:cross-id-accuracy} we plot the performance of our
    method using different binary classification models, as well as the
    performance of Radio Galaxy Zoo, nearest-neighbours, and the perfect and
    random binary classifiers, on the full set of ATLAS~DR1 radio components
    using the pipeline in \autoref{fig:flowchart}. The accuracy
    associated with each classification model and training label set
    averaged across all four quadrants is shown in \aref{app:accuracies}.

    Differences between accuracies across training labels are well within one
    standard deviation computed across the four quadrants, with convolutional
    neural networks on compact objects as the only exception. The spread of
    accuracies is similar for both sets of training labels, with the exception
    of random forests. The balanced accuracies of random forests trained on
    expert labels have a considerably higher spread than those trained on
    Radio Galaxy Zoo labels, likely because of the small size of the expert
    training set --- there are less than half the number of objects in the
    expert-labelled training set than the number of objects in the Radio
    Galaxy Zoo-labelled training set (\autoref{tab:radio-count}).

    Radio Galaxy Zoo-trained methods significantly outperform Radio Galaxy Zoo
    cross-identifications. Additionally, despite poor performance of Radio
    Galaxy Zoo on the cross-identification task, methods trained on these
    cross-identifications still perform comparably to those trained on expert
    labels. This is because incorrect Radio Galaxy Zoo cross-identifications
    can be thought of as a source of noise in the labels which is `averaged out'
    in training. This shows the usefulness of crowdsourced training data, even
    when the data is noisy.

    Our method performs comparably to a nearest-neighbours approach. For
    compact objects, this is to be expected --- indeed, nearest-neighbours
    attains nearly 100 per cent accuracy on the compact test set. Our results
    do not improve on nearest-neighbours for resolved objects. However, our
    method does allow for improvement on nearest-neighbours with a
    sufficiently good binary classifier: a `perfect' binary classifier attains
    nearly 100 per cent accuracy on resolved sources. This shows that our
    method may be useful provided that a good binary classifier can be
    trained. The most obvious place for improvement is in feature selection:
    we use pixels of radio images directly and these are likely not conducive
    to good performance on the candidate classification task. Convolutional
    neural networks, which are able to extract features from images,
    \emph{should} work better, but these require far more training data than
    the other methods we have applied and the small size of ATLAS thus limits their performance.

    We noted in \autoref{sec:labels} that the test set of expert labels,
    derived from the initial ATLAS data release, was less deep than the third
    data release used by Radio Galaxy Zoo and this paper, introducing a source
    of label noise in the testing labels. Specifically, true host galaxies may
    be misidentified as non-host galaxies if the associated radio source was
    below the 5 signal-to-noise limit in ATLAS~DR1 but not in ATLAS~DR3. This
    has the effect of reducing the accuracy for Radio Galaxy Zoo-trained
    classifiers.

    We report the scores predicted by each classifier for each
    SWIRE object in \aref{app:scores} and the predicted
    cross-identification for each ATLAS object in \aref{app:xids}.
    Scores reported for a given object were predicted by binary
    classifiers tested on the quadrant containing that object. The reported scores are not weighted.

    In \autoref{fig:examples} we show 5 resolved sources where the most classifiers disagreed on the correct cross-identification.

\subsection{Application to ATLAS-ELAIS-S1}
  \label{sec:elais}

  \begin{figure*}
  \centering
  \includegraphics[]{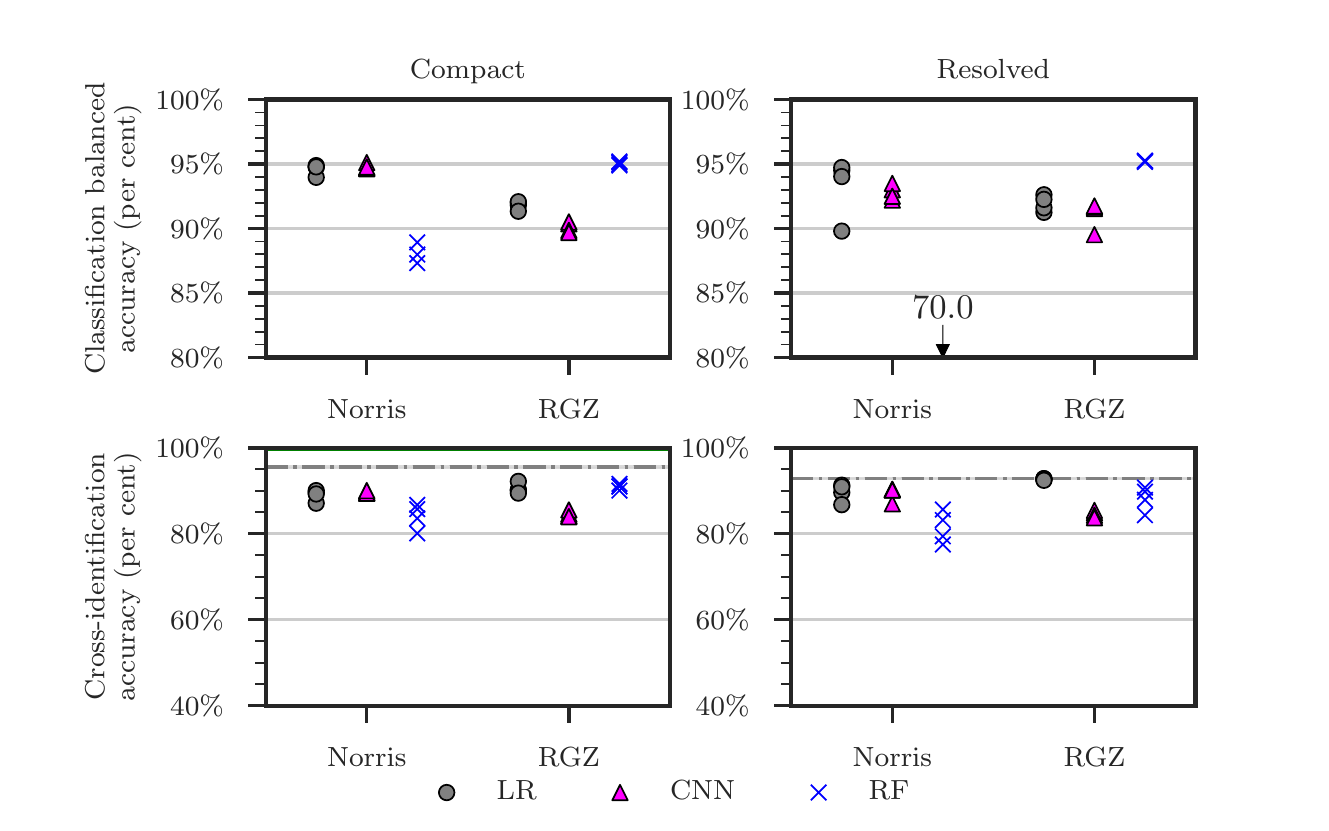}
  \caption{Performance of different classification models trained on CDFS and tested on
  resolved and compact sources in ELAIS-S1. Points represent classification models
  trained on different quadrants of CDFS, with markers, lines and axes as in
  \autoref{fig:ba}. The balanced acccuracy of expert-trained random forest
  binary classifiers falls below the axis and the corresponding mean accuracy is
  shown by an arrow. The estimated best attainable accuracy is almost 100 per cent.
    \label{fig:elais-ba}}
  \end{figure*}

  \begin{figure}
    \centering
    \includegraphics[]{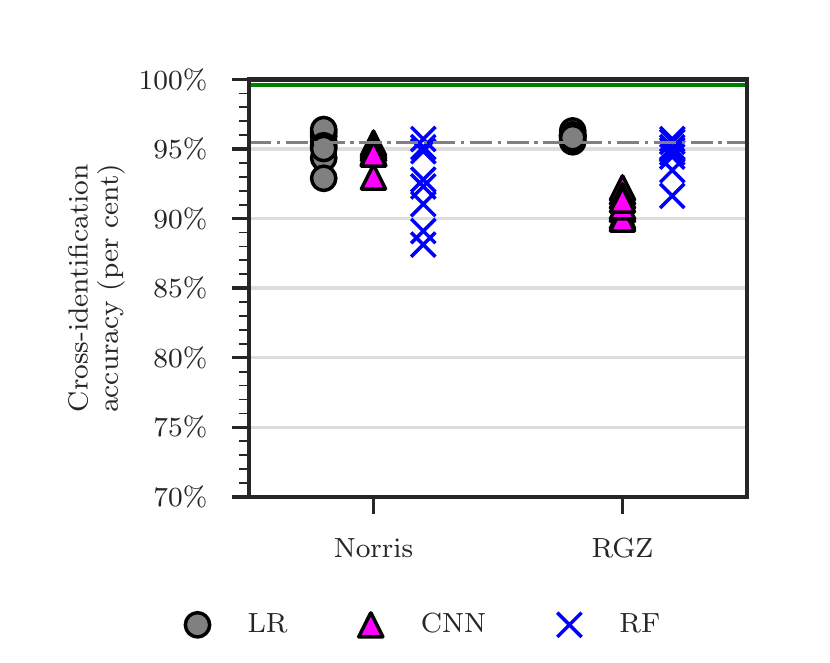}
    \caption{Performance of different classifiers trained on CDFS and tested
    on ELAIS-S1. Markers are as in \autoref{fig:cross-id-accuracy} and
    horizontal lines are as in \autoref{fig:elais-ba}. Note that the pipeline
    shown in \autoref{fig:flowchart} is used here, so compact objects
    are cross-identified in the same way regardless of binary classifier
    model.
      \label{fig:elais-cross-id-accuracy}}
  \end{figure}

  \begin{figure}
    \centering
    \includegraphics[trim={0cm 1cm 0cm 0.5cm}, clip]{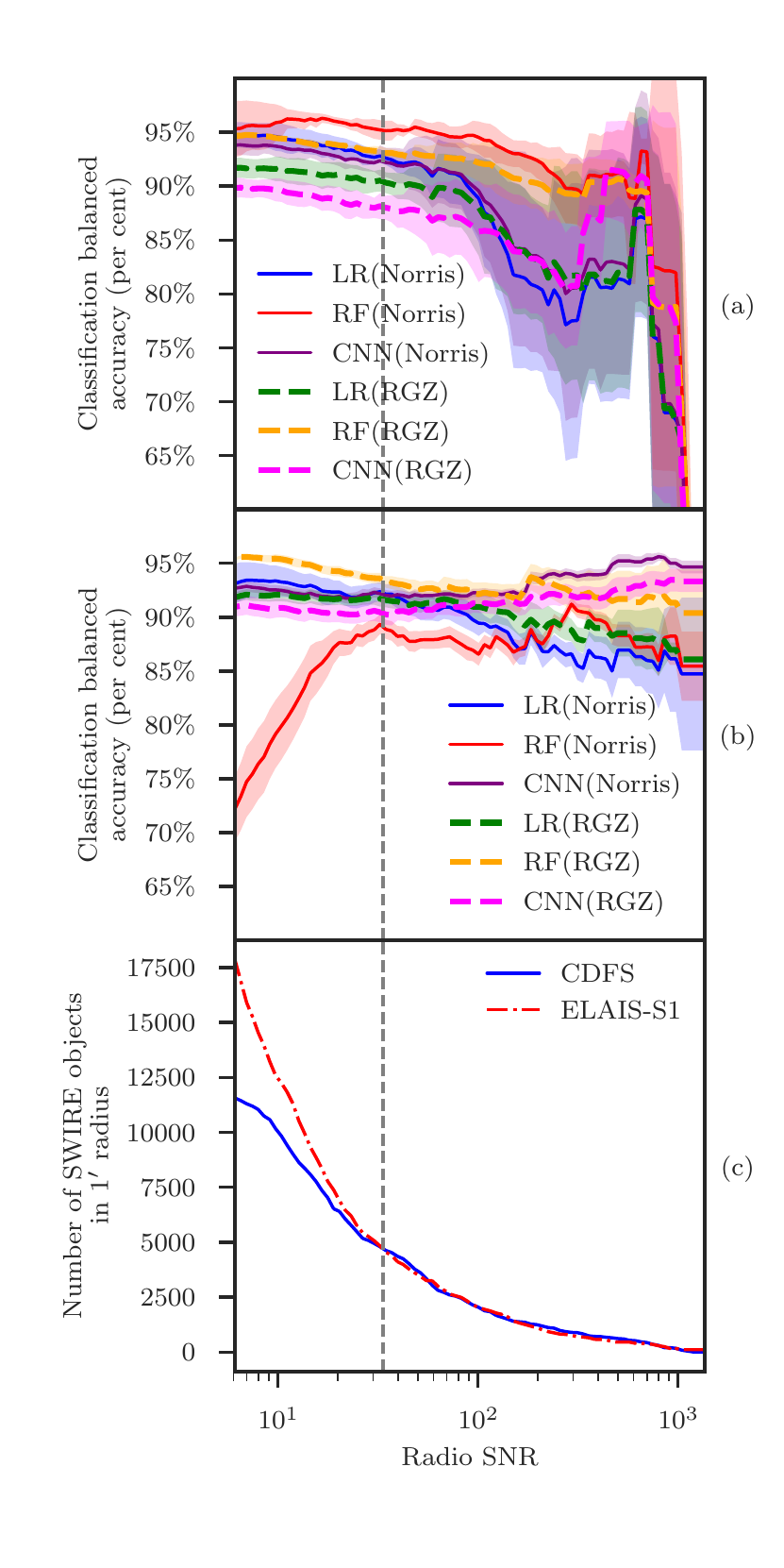}
    \caption{(a) Balanced accuracies of classifiers trained and tested on CDFS
      with different signal-to-noise ratio (SNR) cutoffs for the test set. A
      SWIRE object is included in the test set if it is within $1'$ of a radio
      component with greater SNR than the cutoff. Lines of different colour
      indicate different classifier/training labels combinations, where LR is
      logistic regression, RF is random forests, CNN is convolutional neural
      networks, and Norris and RGZ are the expert and Radio Galaxy Zoo label
      sets respectively. Filled areas represent standard deviations across
      CDFS quadrants. (b) Balanced accuracies of classifiers trained on CDFS
      and tested on ELAIS-S1. (c) A cumulative distribution plot of SWIRE
      objects associated with a radio object with greater SNR than the cutoff.
      The grey dashed line shows the SNR level at which the number of SWIRE
      objects above the cutoff is equal for CDFS and ELAIS-S1. This cutoff level
      is approximately at a SNR of $34$.}
    \label{fig:accuracies-flux}
  \end{figure}

  We applied the method trained on CDFS to perform cross-identification on the
  ELAIS-S1 field. Both CDFS and ELAIS-S1 were imaged by the same radio
  telescope to similar sensitivities and angular resolution for the ATLAS
  survey. We can use the SWIRE cross-identifications made by
  \citet{middelberg08} to derive another set of expert labels, and hence
  determine how accurate our method is. If our method generalises well across
  different parts of the sky, then we expect CDFS-trained classifiers to have
  comparable performance between ELAIS-S1 and CDFS. In \autoref{fig:elais-ba}
  we plot the performance of CDFS-trained classification models on the candidate classification task and
  the performance of our method on the cross-identification task using these models. We also plot
  the cross-identification accuracy of a nearest-neighbours approach\footnote{We cannot
  directly compare our method applied to ELAIS-S1 with Radio Galaxy Zoo, as
  Radio Galaxy Zoo does not include ELAIS-S1.}. In
  \autoref{fig:elais-cross-id-accuracy} we plot the performance of our method
  on the full set of ELAIS-S1 ATLAS~DR1 radio components using the pipeline in
  \autoref{fig:flowchart}. We list the corresponding accuracies in
  \aref{app:accuracies}.

  Cross-identification results from ELAIS-S1 are similar to those for CDFS,
  showing that our method trained on CDFS performs comparably well on
  ELAIS-S1. However, nearest-neighbours outperforms most methods on ELAIS-S1.
  This is likely because there is a much higher percentage of compact objects
  in ELAIS-S1 than in CDFS. The maximum achievable accuracy we have estimated
  for ELAIS-S1 is very close to 100 per cent, so (as for CDFS) a very accurate
  binary classifier would outperform nearest-neighbours.

  One interesting difference between the ATLAS fields is that random forests
  trained on expert labels perform well on CDFS but poorly on ELAIS-S1. This
  is not the case for logistic regression or convolutional neural networks
  trained on expert labels, nor is it the case for random forests trained on
  Radio Galaxy Zoo. We hypothesise that this is because the ELAIS-S1
  cross-identification catalogue \citep{middelberg08} labelled fainter radio
  components than the CDFS cross-identification catalogue \citep{norris06} due
  to noise from the very bright source
  ATCDFS\textunderscore{}J032836.53-284156.0 in CDFS. Classifiers trained on
  CDFS expert labels may thus be biased toward brighter radio components
  compared to ELAIS-S1. Radio Galaxy Zoo uses a preliminary version of the third data release of ATLAS
  \citep{franzen15} and so classifiers trained on the Radio Galaxy Zoo labels
  may be less biased toward brighter sources compared to those trained on the
  expert labels. To test this hypothesis we tested each classification model against
  test sets with a signal-to-noise ratio (SNR) cutoff. A SWIRE object was only
  included in the test set for a given cutoff if it was located within $1'$ of
  a radio component with a SNR above the cutoff. The balanced accuracies for
  each classifier at each cutoff are shown in \autoref{fig:accuracies-flux}(a)
  and (b) and the distribution of test set size for each cutoff is shown in
  \autoref{fig:accuracies-flux}(c). \autoref{fig:accuracies-flux}(c) shows
  that ELAIS-S1 indeed has more faint objects in its test set than the CDFS test set, with the SNR for
  which the two fields reach the same test set size (approximately $34$)
  indicated by the dashed vertical line on each plot. For CDFS, all
  classifiers perform reasonably well across cutoffs, with performance
  dropping as the size of the test set becomes small. For ELAIS-S1, logistic
  regression and convolutional neural networks perform comparably across all
  SNR cutoffs, but random forests do not. While random forests trained on
  Radio Galaxy Zoo labels perform comparably to other classifiers across all
  SNR cutoffs, random forests trained on expert labels show a considerable
  drop in performance below the dashed line.

\section{Discussion}

  Based on the ATLAS sample, our main result is that it is possible
  to cast radio host galaxy cross-identification as a machine learning task
  for which standard methods can be applied. These methods can then be trained
  with a variety of label sets derived from cross-identification catalogues.
  While our methods have not outperformed nearest-neighbours, we have demonstrated that
  for a very accurate binary classifier, good cross-identification results can
  be obtained using our method. Future work could combine multiple catalogues
  or physical priors to boost performance.

  Nearest-neighbours approaches outperform most methods we investigated,
  notably including Radio Galaxy Zoo. This is due to the large number of
  compact or partially-resolved objects in ATLAS. This result shows that for
  compact and partially-resolved objects methods that do not use machine
  learning such as a nearest-neighbours approach or likelihood ratio
  \citep{weston18lrpy} should be preferred to machine learning methods. It
  also shows that ATLAS is not an ideal data set for developing machine
  learning methods like ours. Our use of ATLAS is motivated by its status as a
  pilot survey for EMU, so methods developed for ATLAS should also work for
  EMU. New methods developed should work well with extended radio sources, but
  this goal is almost unsupported by ATLAS as it has very few examples of such
  sources. This makes both training and testing difficult --- there are too
  few extended sources to train on and performance on such a small test set
  may be unreliable. Larger data sets with many extended sources like FIRST
  exist, but these are considerably less deep than and at a different
  resolution to EMU, so there is no reason to expect methods trained on such
  data sets to be applicable to EMU.

  The accuracies of our trained cross-identification methods generally fall
  far below the estimated best possible accuracy attainable using our approach,
  indicated by the green-shaded areas in Figures \ref{fig:cross-id-accuracy} and
  \ref{fig:elais-cross-id-accuracy}. The balanced accuracies attained by our
  binary classifiers indicate that there is significant room for improvement
  in classification. The classification accuracy could be improved by better
  model selection and more training data, particularly for convolutional
  neural networks. There is a huge variety of ways to build a convolutional
  neural network, and we have only investigated one architecture. For an
  exploration of different convolutional neural network architectures applied
  to radio astronomy, see \citet{lukic18compact}. Convolutional neural
  networks generally require more training data than other machine learning
  models and we have only trained our networks on a few hundred sources. We
  would expect performance on the classification task to greatly increase
  with larger training sets.

  Another problem is that of the window size used to select radio features.
  Increasing window size would increase computational expense, but provide
  more information to the models. Results are also highly sensitive to how
  large the window size is compared to the size of the radio source we are
  trying to cross-identify, with large angular sizes requiring large window
  sizes to ensure that the features contain all the information needed to
  localise the host galaxy. An ideal implementation of our method would most
  likely represent a galaxy using radio images taken at multiple window
  sizes, but this is considerably more expensive.

  Larger training sets, better model selection, and larger window sizes would
  improve performance, but only so far: we would still be bounded above by the
  estimated `perfect' classifier accuracy. From this point, the performance
  can only be improved by addressing our broken assumptions. We detailed
  these assumptions in \autoref{sec:limitations}, and we will discuss here how
  our method could be adapted to avoid these assumptions. Our assumption that the host galaxy is contained
  within the search radius could be improved by dynamically choosing the
  search radius, perhaps based on the angular extent of the radio emission, or the
  redshift of candidate hosts. Radio morphology information may allow us to
  select relevant radio data and hence relax the assumption that a $1'$-wide
  radio image represents just one, whole radio source. Finally, our assumption
  that the host galaxy is detefcted in infrared is technically not needed, as
  the sliding-window approach we have employed will still work even if there
  are no detected host galaxies --- instead of classifying candidate hosts,
  simply classify each pixel in the radio image. The downside of removing
  candidate hosts is that we are no longer able to reliably incorporate host
  galaxy information such as colour and redshift, though this could be
  resolved by treating pixels as potentially undetected candidate hosts with
  noisy features.

  We observe that Radio Galaxy Zoo-trained methods perform comparably to
  methods trained on expert labels. This shows that the crowdsourced labels
  from Radio Galaxy Zoo will provide a valuable source of training
  data for future machine learning methods in radio astronomy.

  Compared to nearest-neighbours, cross-identification accuracy on ELAIS-S1 is
  lower than on CDFS. Particularly notable is that our performance on compact
  objects is very low for ELAIS-S1, while it was near-optimal for CDFS. These
  differences may be for a number of reasons. ELAIS-S1 has beam size and noise
  profile different from CDFS (even though both were imaged with the same
  telescope), so it is possible that our methods over-adapted to the beam and
  noise of CDFS. Additionally, CDFS contains a very bright source which may
  have caused artefacts throughout the field that are not present in ELAIS-S1.
  Further work is required to understand the differences between the fields
  and their effect on performance.

  \autoref{fig:accuracies-flux} reveals interesting behaviour of different
  classifier models at different flux cutoffs. Logistic regression and
  convolutional neural networks seem relatively independent of flux, with
  these models performing well on the fainter ELAIS-S1 components even when
  they were trained on the generally brighter components in CDFS. Conversely,
  random forests were sensitive to the changes in flux distribution between
  datasets. This shows that not all models behave similarly on radio data,
  and it is therefore important to investigate multiple models when
  developing machine learning methods for radio astronomy.

  \aref{app:examples} (see \autoref{fig:examples}) shows examples of incorrectly cross-identified
  components in CDFS. On no such component do all classifiers agree.
  This raises the possibility of using the level of disagreement of an
  ensemble of binary classifiers as a measure of the difficulty of cross-identifying a radio component,
  analogous to the consensus level for Radio Galaxy Zoo volunteers.

  Our methods can be easily incorporated into other cross-identification
  methods or used as an extra data source for source detection. For
  example, the scores output by our binary classifiers could be used to
  disambiguate between candidate host
  galaxies selected by model-based algorithms, or used to weight candidate
  host galaxies while a source detector attempts to associate radio
  components. Our method can also be extended using other data sources: for
  example, information from source identification algorithms could be
  incorporated into the feature set of candidate host galaxies.

\section{Summary}

  We presented a machine learning approach for cross-identification of radio
  components with their corresponding infrared host galaxy. Using the CDFS
  field of ATLAS as a training set we trained our
  methods on expert and crowdsourced cross-identification catalogues.
  Applying these methods on both fields of ATLAS, we found that:
  \begin{itemize}
    \item Our method trained on ATLAS observations of CDFS generalised to
    ATLAS observations of ELAIS-S1, demonstrating that training on a single
    patch of sky is a feasible option for training machine learning methods
    for wide-area radio surveys;
    \item Performance was comparable to nearest-neighbours even on resolved
    sources, showing that nearest-neighbours is useful for datasets consisting
    mostly of unresolved sources such as ATLAS and EMU;
    \item Radio Galaxy Zoo-trained models performed comparably to
    expert-trained models and outperformed Radio Galaxy Zoo, showing that
    crowdsourced labels are useful for training machine learning methods for
    cross-identification even when these labels are noisy;
    \item ATLAS does not contain sufficient data to train or test machine
    learning cross-identification methods for extended radio sources. This
    suggests that if machine learning methods are to be used on EMU, a larger
    area of sky will be required for training and testing these methods.
    However, existing surveys like FIRST are likely too different from EMU to expect
    good generalisation.
  \end{itemize}

  While our cross-identification performance is not as high as desired, we
  make no assumptions on the binary classification model used in our methods
  and so we expect the performance to be improved by further experimentation
  and model selection. Our method provides a useful framework for generalising
  cross-identification catalogues to other areas of the sky from the same
  radio survey and can be incorporated into existing methods. We have
  shown that citizen science can provide a useful dataset for training machine
  learning methods in the radio domain.

\section{Acknowledgements}

  This publication has been made possible by the participation of more than
  11~000 volunteers in the Radio Galaxy Zoo project. Their contributions are
  individually acknowledged at \url{http://rgzauthors.galaxyzoo.org}. Parts of
  this research were conducted by the Australian Research Council Centre of
  Excellence for All-sky Astrophysics (CAASTRO), through project number
  CE110001020. Partial support for LR was provided by U.S. National Science
  Foundation grants AST1211595 and 1714205 to the University of Minnesota. HA
  benefitted from grant 980/2016-2017 of Universidad de Guanajuato. We thank
  A.~Tran and the reviewer for their comments on this manuscript.
  Radio Galaxy Zoo makes use of
  data products from the Wide-field Infrared Survey Explorer and the Very
  Large Array. The Wide-field Infrared Survey Explorer is a joint project of
  the University of California, Los Angeles, and the Jet Propulsion
  Laboratory/California Institute of Technology, funded by the National
  Aeronautics and Space Administration. The National Radio Astronomy
  Observatory is a facility of the National Science Foundation operated under
  cooperative agreement by Associated Universities, Inc. The figures
  in this work made use of Astropy, a community-developed core Python package
  for Astronomy \citep{astropy}. The Australia Telescope Compact Array is part
  of the Australia Telescope, which is funded by the Commonwealth of Australia
  for operation as a National Facility managed by CSIRO.

\bibliographystyle{mnras}
\bibliography{rgz-cdfs-ms}

\begin{thebibliography}{}
\makeatletter
\relax
\def\mn@urlcharsother{\let\do\@makeother \do\$\do\&\do\#\do\^\do\_\do\%\do\~}
\def\mn@doi{\begingroup\mn@urlcharsother \@ifnextchar [ {\mn@doi@}
  {\mn@doi@[]}}
\def\mn@doi@[#1]#2{\def\@tempa{#1}\ifx\@tempa\@empty \href
  {http://dx.doi.org/#2} {doi:#2}\else \href {http://dx.doi.org/#2} {#1}\fi
  \endgroup}
\def\mn@eprint#1#2{\mn@eprint@#1:#2::\@nil}
\def\mn@eprint@arXiv#1{\href {http://arxiv.org/abs/#1} {{\tt arXiv:#1}}}
\def\mn@eprint@dblp#1{\href {http://dblp.uni-trier.de/rec/bibtex/#1.xml}
  {dblp:#1}}
\def\mn@eprint@#1:#2:#3:#4\@nil{\def\@tempa {#1}\def\@tempb {#2}\def\@tempc
  {#3}\ifx \@tempc \@empty \let \@tempc \@tempb \let \@tempb \@tempa \fi \ifx
  \@tempb \@empty \def\@tempb {arXiv}\fi \@ifundefined
  {mn@eprint@\@tempb}{\@tempb:\@tempc}{\expandafter \expandafter \csname
  mn@eprint@\@tempb\endcsname \expandafter{\@tempc}}}

\bibitem[\protect\citeauthoryear{{Aniyan} \& {Thorat}}{{Aniyan} \&
  {Thorat}}{2017}]{aniyan17cnn}
{Aniyan} A.~K.,  {Thorat} K.,  2017, \mn@doi [\apjs]
  {10.3847/1538-4365/aa7333}, \href
  {http://adsabs.harvard.edu/abs/2017ApJS..230...20A} {230, 20}

\bibitem[\protect\citeauthoryear{{Astropy Collaboration} et~al.,}{{Astropy
  Collaboration} et~al.}{2013}]{astropy}
{Astropy Collaboration} et~al., 2013, \mn@doi [\aap]
  {10.1051/0004-6361/201322068}, \href
  {http://adsabs.harvard.edu/abs/2013A%26A...558A..33A} {558, A33}

\bibitem[\protect\citeauthoryear{{Banfield} et~al.,}{{Banfield}
  et~al.}{2015}]{banfield15}
{Banfield} J.~K.,  et~al., 2015, \mn@doi [\mnras] {10.1093/mnras/stv1688},
  \href {http://adsabs.harvard.edu/abs/2015MNRAS.453.2326B} {453, 2326}

\bibitem[\protect\citeauthoryear{{Bertin} \& {Arnouts}}{{Bertin} \&
  {Arnouts}}{1996}]{bertin96sextractor}
{Bertin} E.,  {Arnouts} S.,  1996, \mn@doi [\aaps] {10.1051/aas:1996164}, \href
  {http://adsabs.harvard.edu/abs/1996A%26AS..117..393B} {117, 393}

\bibitem[\protect\citeauthoryear{Bishop}{Bishop}{2006}]{bishop06ml}
Bishop C.~M.,  2006, Pattern recognition and machine learning.
Springer

\bibitem[\protect\citeauthoryear{Breiman}{Breiman}{2001}]{breiman01random-forest}
Breiman L.,  2001, Machine Learning, 45, 5

\bibitem[\protect\citeauthoryear{Chollet et~al.}{Chollet
  et~al.}{2015}]{chollet15keras}
Chollet F.,  et~al., 2015, Keras, \url{https://github.com/fchollet/keras}

\bibitem[\protect\citeauthoryear{{Collier} et~al.,}{{Collier}
  et~al.}{2014}]{collier14irfaint}
{Collier} J.~D.,  et~al., 2014, \mn@doi [\mnras] {10.1093/mnras/stt2485}, \href
  {http://adsabs.harvard.edu/abs/2014MNRAS.439..545C} {439, 545}

\bibitem[\protect\citeauthoryear{Cutri et~al.,}{Cutri
  et~al.}{2013}]{cutri2013wiseexplanatory}
Cutri R.,  et~al., 2013, Explanatory Supplement to the AllWISE Data Release
  Products, by RM Cutri et al.

\bibitem[\protect\citeauthoryear{{Dieleman}, {Willett}  \& {Dambre}}{{Dieleman}
  et~al.}{2015}]{dieleman15cnn}
{Dieleman} S.,  {Willett} K.~W.,   {Dambre} J.,  2015, \mn@doi [\mnras]
  {10.1093/mnras/stv632}, \href
  {http://adsabs.harvard.edu/abs/2015MNRAS.450.1441D} {450, 1441}

\bibitem[\protect\citeauthoryear{Fan, Budav\'ari, Norris  \& Hopkins}{Fan
  et~al.}{2015}]{fan15}
Fan D.,  Budav\'ari T.,  Norris R.~P.,   Hopkins A.~M.,  2015, \mn@doi [\mnras]
  {10.1093/mnras/stv994}, 451, 1299

\bibitem[\protect\citeauthoryear{{Franzen} et~al.,}{{Franzen}
  et~al.}{2015}]{franzen15}
{Franzen} T.~M.~O.,  et~al., 2015, \mn@doi [\mnras] {10.1093/mnras/stv1866},
  \href {http://adsabs.harvard.edu/abs/2015MNRAS.453.4020F} {453, 4020}

\bibitem[\protect\citeauthoryear{{Gendre} \& {Wall}}{{Gendre} \&
  {Wall}}{2008}]{Gendre2008}
{Gendre} M.~A.,  {Wall} J.~V.,  2008, \mn@doi [\mnras]
  {10.1111/j.1365-2966.2008.13792.x}, \href
  {http://adsabs.harvard.edu/abs/2008MNRAS.390..819G} {390, 819}

\bibitem[\protect\citeauthoryear{Grant}{Grant}{2011}]{grant11polarised}
Grant J.~K.,  2011, PhD thesis, University of Calgary

\bibitem[\protect\citeauthoryear{{Grant}, {Taylor}, {Stil}, {Landecker},
  {Kothes}, {Ransom}  \& {Scott}}{{Grant} et~al.}{2010}]{Grant2010}
{Grant} J.~K.,  {Taylor} A.~R.,  {Stil} J.~M.,  {Landecker} T.~L.,  {Kothes}
  R.,  {Ransom} R.~R.,   {Scott} D.,  2010, \mn@doi [\apj]
  {10.1088/0004-637X/714/2/1689}, \href
  {http://adsabs.harvard.edu/abs/2010ApJ...714.1689G} {714, 1689}

\bibitem[\protect\citeauthoryear{{Johnston} et~al.,}{{Johnston}
  et~al.}{2007}]{johnston07}
{Johnston} S.,  et~al., 2007, \mn@doi [\pasa] {10.1071/AS07033}, \href
  {http://adsabs.harvard.edu/abs/2007PASA...24..174J} {24, 174}

\bibitem[\protect\citeauthoryear{LeCun, Bottou, Bengio  \& Haffner}{LeCun
  et~al.}{1998}]{lecun98}
LeCun Y.,  Bottou L.,  Bengio Y.,   Haffner P.,  1998, Proceedings of the IEEE,
  86, 2278

\bibitem[\protect\citeauthoryear{{Lintott} et~al.,}{{Lintott}
  et~al.}{2008}]{lintott08}
{Lintott} C.~J.,  et~al., 2008, \mn@doi [\mnras]
  {10.1111/j.1365-2966.2008.13689.x}, \href
  {http://adsabs.harvard.edu/abs/2008MNRAS.389.1179L} {389, 1179}

\bibitem[\protect\citeauthoryear{{Lonsdale} et~al.,}{{Lonsdale}
  et~al.}{2003}]{lonsdale03swire}
{Lonsdale} C.~J.,  et~al., 2003, \mn@doi [\pasp] {10.1086/376850}, \href
  {http://adsabs.harvard.edu/abs/2003PASP..115..897L} {115, 897}

\bibitem[\protect\citeauthoryear{{Lukic}, {Br{\"u}ggen}, {Banfield}, {Wong},
  {Rudnick}, {Norris}  \& {Simmons}}{{Lukic} et~al.}{2018}]{lukic18compact}
{Lukic} V.,  {Br{\"u}ggen} M.,  {Banfield} J.~K.,  {Wong} O.~I.,  {Rudnick} L.,
   {Norris} R.~P.,   {Simmons} B.,  2018, \mn@doi [\mnras]
  {10.1093/mnras/sty163}, \href
  {http://adsabs.harvard.edu/abs/2018MNRAS.476..246L} {476, 246}

\bibitem[\protect\citeauthoryear{Marshall, Lintott  \& Fletcher}{Marshall
  et~al.}{2015}]{marshall15citizenscience}
Marshall P.~J.,  Lintott C.~J.,   Fletcher L.~N.,  2015, \mn@doi [Annual Review
  of Astronomy and Astrophysics] {10.1146/annurev-astro-081913-035959}, 53, 247

\bibitem[\protect\citeauthoryear{Menon, Van~Rooyen, Ong  \& Williamson}{Menon
  et~al.}{2015}]{menon15cpe}
Menon A.~K.,  Van~Rooyen B.,  Ong C.~S.,   Williamson R.~C.,  2015, in
  Proceedings of the 32Nd International Conference on International Conference
  on Machine Learning - Volume 37. ICML'15.
JMLR.org, pp 125--134, \url {http://dl.acm.org/citation.cfm?id=3045118.3045133}

\bibitem[\protect\citeauthoryear{{Middelberg} et~al.,}{{Middelberg}
  et~al.}{2008}]{middelberg08}
{Middelberg} E.,  et~al., 2008, \mn@doi [\aj] {10.1088/0004-6256/135/4/1276},
  \href {http://adsabs.harvard.edu/abs/2008AJ....135.1276M} {135, 1276}

\bibitem[\protect\citeauthoryear{{Norris}}{{Norris}}{2017a}]{norris17surveys}
{Norris} R.~P.,  2017a, \mn@doi [Nature Astronomy] {10.1038/s41550-017-0233-y},
  \href {http://adsabs.harvard.edu/abs/2017NatAs...1..671N} {1, 671}

\bibitem[\protect\citeauthoryear{{Norris}}{{Norris}}{2017b}]{norris17unexpected}
{Norris} R.~P.,  2017b, \mn@doi [\pasa] {10.1017/pasa.2016.63}, \href
  {http://adsabs.harvard.edu/abs/2017PASA...34....7N} {34, e007}

\bibitem[\protect\citeauthoryear{{Norris} et~al.,}{{Norris}
  et~al.}{2006}]{norris06}
{Norris} R.~P.,  et~al., 2006, \mn@doi [\aj] {10.1086/508275}, \href
  {http://adsabs.harvard.edu/abs/2006AJ....132.2409N} {132, 2409}

\bibitem[\protect\citeauthoryear{{Norris} et~al.,}{{Norris}
  et~al.}{2011}]{norris11}
{Norris} R.~P.,  et~al., 2011, \mn@doi [\pasa] {10.1071/AS11021}, \href
  {http://adsabs.harvard.edu/abs/2011PASA...28..215N} {28, 215}

\bibitem[\protect\citeauthoryear{Pedregosa et~al.,}{Pedregosa
  et~al.}{2011}]{pedregosa11sklearn}
Pedregosa F.,  et~al., 2011, J. Mach. Learn. Res., 12, 2825

\bibitem[\protect\citeauthoryear{{Proctor}}{{Proctor}}{2006}]{proctor06}
{Proctor} D.~D.,  2006, \mn@doi [\apjs] {10.1086/504801}, \href
  {http://adsabs.harvard.edu/abs/2006ApJS..165...95P} {165, 95}

\bibitem[\protect\citeauthoryear{{Richter}}{{Richter}}{1975}]{richter75likelihood}
{Richter} G.~A.,  1975, \mn@doi [Astronomische Nachrichten]
  {10.1002/asna.19752960203}, \href
  {http://adsabs.harvard.edu/abs/1975AN....296...65R} {296, 65}

\bibitem[\protect\citeauthoryear{Rowley, Baluja  \& Kanade}{Rowley
  et~al.}{1996}]{rowley1996facedetection}
Rowley H.~A.,  Baluja S.,   Kanade T.,  1996, in Advances in Neural Information
  Processing Systems. p.~875

\bibitem[\protect\citeauthoryear{{Sajina}, {Lacy}  \& {Scott}}{{Sajina}
  et~al.}{2005}]{Sajina2005}
{Sajina} A.,  {Lacy} M.,   {Scott} D.,  2005, \mn@doi [\apj] {10.1086/426536},
  \href {http://adsabs.harvard.edu/abs/2005ApJ...621..256S} {621, 256}

\bibitem[\protect\citeauthoryear{Surace et~al.,}{Surace
  et~al.}{2005}]{surace05swire}
Surace J.,  et~al., 2005, Spitzer Science Centre, California Institute of
  Technology, Pasadena, CA

\bibitem[\protect\citeauthoryear{{Taylor} et~al.,}{{Taylor}
  et~al.}{2007}]{Taylor2007}
{Taylor} A.~R.,  et~al., 2007, \mn@doi [\apj] {10.1086/519786}, \href
  {http://adsabs.harvard.edu/abs/2007ApJ...666..201T} {666, 201}

\bibitem[\protect\citeauthoryear{{Verheijen}, {Oosterloo}, {van Cappellen},
  {Bakker}, {Ivashina}  \& {van der Hulst}}{{Verheijen}
  et~al.}{2008}]{verheijen08}
{Verheijen} M.~A.~W.,  {Oosterloo} T.~A.,  {van Cappellen} W.~A.,  {Bakker} L.,
   {Ivashina} M.~V.,   {van der Hulst} J.~M.,  2008, in AIP Conf. Ser. 1035,
  The Evolution of Galaxies Through the Neutral Hydrogen Window. pp 265--271

\bibitem[\protect\citeauthoryear{{Weston}, {Seymour}, {Gulyaev}, {Norris},
  {Banfield}, {Vaccari}, {Hopkins}  \& {Franzen}}{{Weston}
  et~al.}{2018}]{weston18lrpy}
{Weston} S.~D.,  {Seymour} N.,  {Gulyaev} S.,  {Norris} R.~P.,  {Banfield} J.,
  {Vaccari} M.,  {Hopkins} A.~M.,   {Franzen} T.~M.~O.,  2018, \mn@doi [\mnras]
  {10.1093/mnras/stx2562}, \href
  {http://adsabs.harvard.edu/abs/2018MNRAS.473.4523W} {473, 4523}

\bibitem[\protect\citeauthoryear{{White}, {Becker}, {Helfand}  \&
  {Gregg}}{{White} et~al.}{1997}]{white97first}
{White} R.~L.,  {Becker} R.~H.,  {Helfand} D.~J.,   {Gregg} M.~D.,  1997,
  \mn@doi [\apj] {10.1086/303564}, \href
  {http://adsabs.harvard.edu/abs/1997ApJ...475..479W} {475, 479}

\makeatother
\end{thebibliography}

\appendix

\section{Classification models}\label{app:models}

  We use three different models for binary classification: logistic
  regression, convolutional neural networks, and random forests.

  \subsection{Logistic Regression}
  \label{sec:logistic-regression}
    Logistic regression is linear in the feature space and outputs the
    probability that the input has a positive label. The model is
    \citep{bishop06ml}:

    \begin{equation}
        f(\vec x) = \sigma(\vec w^T \vec x + b) \,\,\,\,,
    \end{equation}
    where $\vec w \in \mathbb{R}^D$ is a vector of parameters, $b \in \mathbb{R}$ is a bias term, $\vec x \in \mathbb{R}^D$ is the feature vector representation of a candidate host, and $\sigma : \mathbb{R} \to \mathbb{R}$ is the logistic sigmoid function: \begin{equation}
        \sigma(a) = (1 + \mathrm{exp}(-a))^{-1}\,\,\,\,.
    \end{equation}%
    The logistic regression model is fully differentiable, and the parameters
    $\vec w$ can therefore be learned using gradient-based optimisation
    methods. We used the \texttt{scikit-learn} \citep{pedregosa11sklearn}
    implementation of logistic regression with balanced classes.

  \subsection{Convolutional neural networks}
  \label{sec:convolutional-neural-networks}

    Convolutional neural networks (CNN) are a biologically-inspired prediction
    model for prediction with image inputs. The input image is convolved with
    a number of filters to produce output images called feature maps. These
    feature maps can then be convolved again with other filters on subsequent
    layers, producing a network of convolutions. The whole network is
    differentiable with respect to the values of the filters and the filters
    can be learned using gradient-based optimisation methods. The final layer
    of the network is logistic regression, with the convolved outputs as input
    features. For more detail, see \citet[subsection II.A][]{lecun98}. We use
    \textsc{Keras} \citep{chollet15keras} to implement our CNN, accounting for
    class imbalance by reweighting the classes.

    \begin{figure}
      \includegraphics[width=\linewidth]{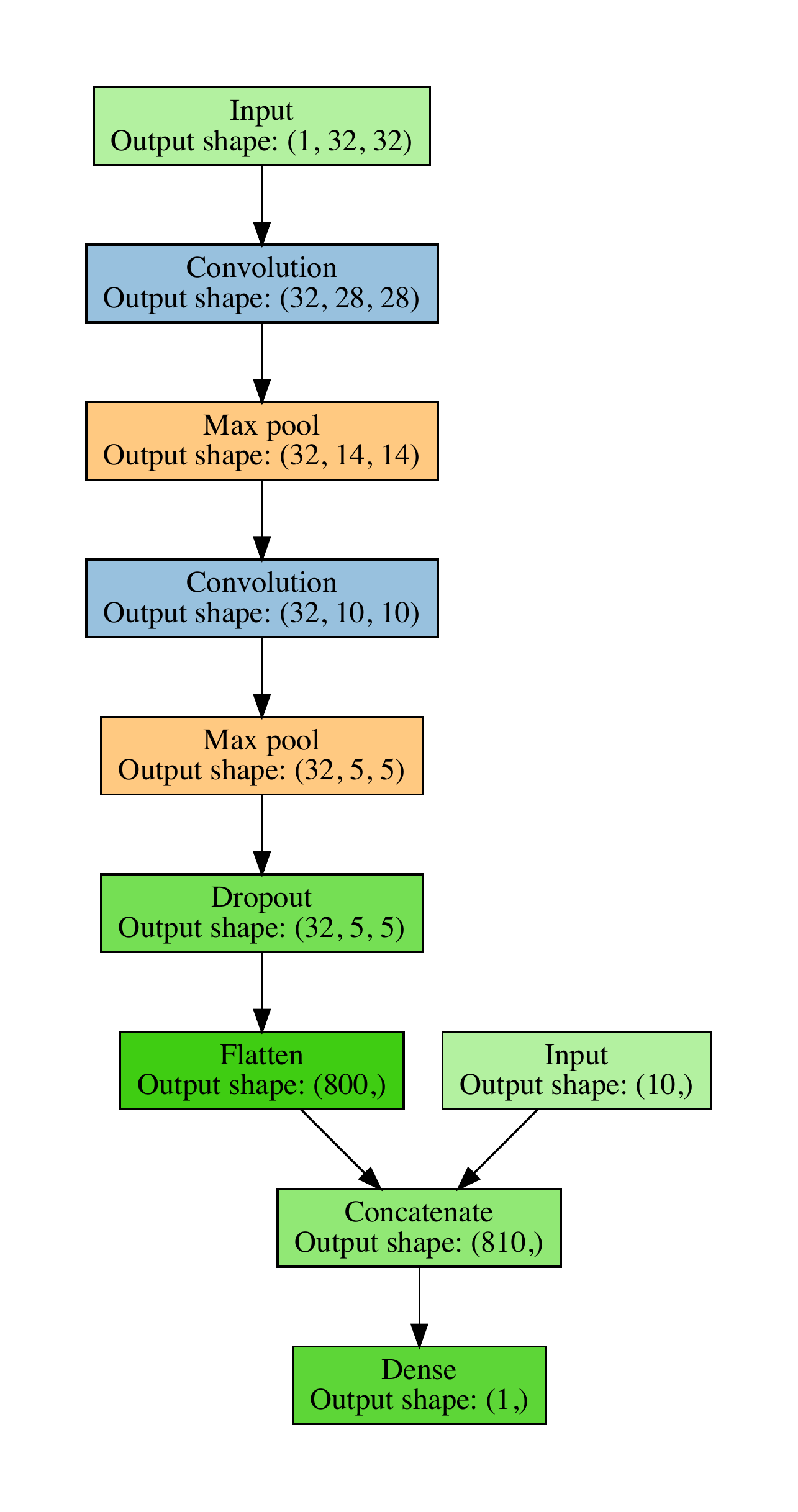}
      \caption{Architecture of our CNN. Parenthesised numbers indicate
      the size of output layers as a tuple (width, height, depth). The
      concatenate layer flattens the output of the previous layer and adds the
      10 features derived from the candidate host in SWIRE, i.e. the flux
      ratios, stellarity indices, and distance. The dropout layer randomly
      sets $25\%$ of its inputs to zero during training to prevent
      overfitting. Diagram based on \url{ https://github.com/dnouri/nolearn}.}
      \label{fig:cnn}
    \end{figure}

    CNNs have recently produced good results on large image-based datasets in
    astronomy \citep[e.g.]{lukic18compact, dieleman15cnn}. We employ
    only a simple CNN model in this paper as a proof of concept that CNNs may
    be used for class probability prediction on radio images. The model
    architecture we use is shown in \autoref{fig:cnn}.

  \subsection{Random Forests}
  \label{sec:random-forests}

    Random forests are an ensemble of decision
    trees~\citep{breiman01random-forest}. They consider multiple subsamples of
    the training set, where each subsample is sampled with replacement from
    the training set. For each subsample a decision tree classifier is
    constructed by repeatedly making axis-parallel splits based on individual
    features. In a random forest the split decision is taken based on a random
    subset of features. To classify a new data point, the random forest takes
    the weighted average of all classifications produced by each decision
    tree. We used the \texttt{scikit-learn} \citep{pedregosa11sklearn}
    implementation of random forests with 10 trees, the information entropy
    split criterion, a minimum leaf size of 45 and balanced classes.

\section{Accuracy tables}\label{app:accuracies}
  
  This section contains tables of accuracy for our method applied to CDFS and
  ELAIS-S1. In \autoref{tab:cdfs-ba} and \autoref{tab:elais-ba} we list the
  balanced accuracies of classifiers on the cross-identification task for CDFS
  and ELAIS-S1 respectively, averaged over each set of training quadrants. In
  \autoref{tab:cdfs-acc} and \autoref{tab:elais-acc} we list the balanced
  accuracies of classifiers on the cross-identification task for CDFS and
  ELAIS-S1 respectively, averaged over each set of training quadrants.

  \begin{table*}
    \caption{Balanced accuracies for different binary classification models trained and tested on SWIRE objects in CDFS.
    The `Labeller' column states what set of training labels
    were used to train the classifier, and the `Classifier' column states what
    classification model was used. `CNN' is a convolutional neural network,
    `LR' is logistic regression and `RF' is random forests. Accuracies are evaluated against the expert
    label set derived from \citet{norris06}. The standard deviation of balanced accuracies evaluated across the four quadrants of
    CDFS (\autoref{fig:quadrants}) is also shown. The `compact' set refers to SWIRE
    objects within $1'$ of a compact radio component, the `resolved' set refers to
    SWIRE objects within $1'$ of a resolved radio component, and `all' is the union of these sets.}
    \label{tab:cdfs-ba}
    \begin{tabular}{ccccc}
    \hline
    Labeller & Classifier & Mean `Compact' accuracy & Mean `Resolved' accuracy & Mean `All' accuracy\\
     &  & (per cent) & (per cent) & (per cent)\\
    \hline
    Norris & LR & $91.5 \pm 1.0$ & $93.2 \pm 1.0$ & $93.0 \pm 1.2$\\
    Norris & CNN & $92.6 \pm 0.7$ & $91.2 \pm 0.5$ & $92.0 \pm 0.6$\\
    Norris & RF & $96.7 \pm 1.5$ & $91.0 \pm 4.5$ & $96.0 \pm 2.5$\\
    RGZ & LR & $89.5 \pm 0.8$ & $90.5 \pm 1.7$ & $90.2 \pm 0.8$\\
    RGZ & CNN & $89.4 \pm 0.6$ & $89.6 \pm 1.3$ & $89.4 \pm 0.5$\\
    RGZ & RF & $94.5 \pm 0.2$ & $95.8 \pm 0.4$ & $94.7 \pm 0.3$\\
    \hline
    \end{tabular}
  \end{table*}

  \begin{table*}
    \caption{Balanced accuracies for different binary classification models trained on SWIRE objects
    in CDFS and tested on SWIRE objects in ELAIS-S1. Columns and abbreviations are as in \autoref{tab:cdfs-ba}. Accuracies are evaluated against the expert
    label set derived from \citet{middelberg08}. The standard deviations of balanced accuracies of models trained on the four subsets of
    CDFS (\autoref{fig:quadrants}) are also shown.}
    \label{tab:elais-ba}
    \begin{tabular}{ccccc}
      \hline
      Labeller & Classifier & Mean `Compact' accuracy & Mean `Resolved' accuracy & Mean `All' accuracy\\
        &  & (per cent) & (per cent) & (per cent)\\
      \hline
      Norris & LR & $94.6 \pm 0.4$ & $93.3 \pm 2.0$ & $95.3 \pm 0.1$\\
       & CNN & $94.8 \pm 0.2$ & $92.8 \pm 0.5$ & $94.4 \pm 0.2$\\
       & RF & $85.9 \pm 3.8$ & $70.0 \pm 2.8$ & $86.6 \pm 3.2$\\
      RGZ & LR & $91.8 \pm 0.3$ & $91.9 \pm 0.5$ & $92.0 \pm 0.2$\\
       & CNN & $90.1 \pm 0.3$ & $91.1 \pm 0.9$ & $90.2 \pm 0.3$\\
       & RF & $95.1 \pm 0.1$ & $95.2 \pm 0.0$ & $95.2 \pm 0.3$\\
      \hline
    \end{tabular}
  \end{table*}

  \begin{table*}
    \caption{Cross-identification accuracies for different classification
    models on CDFS. The `Labeller' column states what set of training labels
    were used to train the method, and the `Classifier' column states what
    classification model was used. `CNN' is a convolutional neural network,
    `LR' is logistic regression, `RF' is random forests, and `Labels' is the
    accuracy of the label set itself. `Perfect' indicates that the true labels
    of the test set were used and hence represents an upper bound on
    cross-identification accuracy with our method. `NN' is a
    nearest-neighbours approach. Accuracies are evaluated against the expert
    label set, so `Norris' labels are 100 per cent accurate by definition. The
    standard deviation of accuracies evaluated across the four quadrants of
    CDFS (\autoref{fig:quadrants}) is also shown.}
    \label{tab:cdfs-acc}
    \begin{tabular}{ccccc}
      \hline
      Labeller & Classifier & Mean `Compact' accuracy & Mean `Resolved' accuracy & Mean `All' accuracy\\
       &  & (per cent) & (per cent) & (per cent)\\
      \hline
      --- & NN & $97.2 \pm 1.7$ & $75.7 \pm 7.9$ & $93.4 \pm 0.8$\\
      --- & Random & $97.9 \pm 2.2$ & $22.3 \pm 9.2$ & $83.2 \pm 4.7$\\
      Norris & Labels & $100.0 \pm 0.0$ & $100.0 \pm 0.0$ & $100.0 \pm 0.0$\\
             & Perfect & $97.9 \pm 2.2$ & $99.0 \pm 1.8$ & $98.1 \pm 1.7$\\
             & LR & $97.3 \pm 0.5$ & $76.0 \pm 3.2$ & $93.7 \pm 1.8$\\
             & CNN & $96.6 \pm 0.9$ & $74.3 \pm 12.3$ & $93.5 \pm 0.5$\\
             & RF & $96.1 \pm 1.4$ & $75.8 \pm 6.7$ & $93.8 \pm 2.0$\\
      RGZ & Labels & $53.1 \pm 8.5$ & $56.7 \pm 5.9$ & $54.4 \pm 5.9$\\
          & LR & $97.3 \pm 1.9$ & $74.5 \pm 5.1$ & $93.6 \pm 1.7$\\
          & CNN & $85.4 \pm 2.6$ & $68.1 \pm 9.2$ & $92.4 \pm 1.1$\\
          & RF & $97.5 \pm 0.9$ & $74.3 \pm 7.9$ & $93.7 \pm 1.5$\\
      \hline
    \end{tabular}
  \end{table*}

  \begin{table*}
    \caption{Cross-identification accuracies for different classification
    models on ELAIS-S1. Columns and abbreviations are as in
    \autoref{tab:cdfs-acc}. Accuracies are evaluated against the expert label
    set derived from \citet{middelberg08} cross-identifications. The standard
    deviation of accuracies evaluated across models trained on the four
    quadrants of CDFS (\autoref{fig:quadrants}) is also shown.}
    \label{tab:elais-acc}
    \begin{tabular}{ccccc}
      \hline
      Labeller & Classifier & Mean `Compact' accuracy & Mean `Resolved' accuracy & Mean `All' accuracy\\
       &  & (per cent) & (per cent) & (per cent)\\
      \hline
      --- & NN & $95.5 \pm 0.0$ & $92.8 \pm 0.0$ & $95.5 \pm 0.0$\\
      --- & Random & $61.9 \pm 1.1$ & $26.6 \pm 2.1$ & $61.9 \pm 1.1$\\
      Middelberg & Perfect & $99.6 \pm 0.0$ & $99.8 \pm 0.0$ & $99.6 \pm 0.0$\\
      Norris & LR & $89.0 \pm 1.1$ & $89.7 \pm 1.8$ & $94.4 \pm 0.9$\\
             & CNN & $89.7 \pm 0.3$ & $89.4 \pm 1.4$ & $94.3 \pm 0.7$\\
             & RF & $83.8 \pm 5.6$ & $82.3 \pm 4.1$ & $90.6 \pm 2.1$\\
      RGZ & LR & $90.5 \pm 1.0$ & $92.7 \pm 0.2$ & $95.9 \pm 0.1$\\
          & CNN & $84.6 \pm 0.6$ & $84.6 \pm 0.6$ & $91.8 \pm 0.3$\\
          & RF & $91.3 \pm 1.0$ & $90.3 \pm 2.4$ & $94.7 \pm 1.2$\\
      \hline
    \end{tabular}
  \end{table*}

\section{SWIRE object scores}\label{app:scores}
  
  This section contains scores predicted by our binary classifiers for each
  SWIRE object within $1'$ of a radio component in CDFS and ELAIS-S1. Scores
  for SWIRE~CDFS objects are shown in \autoref{tab:cdfs-scores} and scores for
  SWIRE~ELAIS-S1 are shown in \autoref{tab:elais-scores}. For CDFS, the score
  for an object in a quadrant is predicted by binary classifiers trained on
  all other quadrants. For ELAIS-S1, we show the scores predicted by binary
  classifiers trained on each CDFS quadrant. Note that these scores have
  \emph{not} been weighted by Gaussians.

  The columns of the score tables are defined as follows:
  \begin{itemize}
    \item \emph{SWIRE} --- SWIRE designation for candidate host galaxy.
    \item \emph{RA} --- Right ascension (J2000).
    \item \emph{Dec} --- Declination (J2000).
    \item \emph{Expert host} --- Whether the candidate host galaxy is a host galaxy according to \citet{norris06} or \citet{middelberg08} cross-identifications of CDFS and ELAIS-S1 respectively.
    \item \emph{RGZ host} --- Whether the candidate host galaxy is a host galaxy according to Radio Galaxy Zoo cross-identifications (Wong et al. in prep). This is always "no" for ELAIS-S1 objects.
    \item \emph{$C$($L$ / $D$)} --- Score assigned by binary classifier $C$ trained on label set $L$ of $D$ candidate host galaxies. $C$ may be `CNN', `LR' or `RF' for CNN, logistic regression or random forests respectively. $L$ may be `Norris' or `RGZ' for expert and Radio Galaxy Zoo labels respectively. $D$ may be `All', `Compact' or `Resolved' for each respective subset defined in \autoref{sec:experimental-setup}.
  \end{itemize}

  \begin{table*}
    \caption{Scores output by our trained classifiers for SWIRE~CDFS candidate host galaxies. Columns are defined in \autoref{app:scores}. Full table electronic.}
    \label{tab:cdfs-scores}
    \begin{tabular}{ccccccccccccccccccccccc}
      \hline
SWIRE & RA & Dec & Expert host & RGZ host & CNN(Norris / All) & CNN(Norris / Compact) & CNN(Norris / Resolved) & CNN(RGZ / All) & CNN(RGZ / Compact) & CNN(RGZ / Resolved) & LR(Norris / All) & LR(Norris / Compact) & LR(Norris / Resolved) & LR(RGZ / All) & LR(RGZ / Compact) & LR(RGZ / Resolved) & RF(Norris / All) & RF(Norris / Compact) & RF(Norris / Resolved) & RF(RGZ / All) & RF(RGZ / Compact) & RF(RGZ / Resolved) \\
      \hline
J032603.15-284708.5 & 51.5132 & -28.7857 & yes & no & 0.5838 & 0.4697 & 0.4848 & 0.3754 & 0.3881 & 0.3404 & 0.2489 & 0.0009 & 0.1557 & 0.2939 & 0.0007 & 0.1174 & 0.8922 & 0.8018 & 0.8732 & 0.7167 & 0.6599 & 0.7801 \\
J032603.39-284010.1 & 51.5142 & -28.6695 & no & no & 0.0373 & 0.5814 & 0.4878 & 0.7896 & 0.7616 & 0.4668 & 0.0183 & 0.1646 & 0.1480 & 0.7637 & 0.7065 & 0.6070 & 0.0000 & 0.0000 & 0.0000 & 0.1629 & 0.0519 & 0.1275 \\
J032603.44-284210.1 & 51.5144 & -28.7028 & no & no & 0.0232 & 0.4891 & 0.5101 & 0.4319 & 0.4298 & 0.3474 & 0.0155 & 0.0164 & 0.0815 & 0.3714 & 0.5626 & 0.2488 & 0.0000 & 0.0734 & 0.0000 & 0.1315 & 0.2116 & 0.4150 \\
J032603.44-284222.2 & 51.5143 & -28.7062 & no & no & 0.0006 & 0.4164 & 0.5216 & 0.0400 & 0.0444 & 0.0276 & 0.0005 & 0.0006 & 0.0175 & 0.0460 & 0.0810 & 0.0299 & 0.2656 & 0.1418 & 0.0000 & 0.7631 & 0.8166 & 0.5378 \\
J032603.45-284748.4 & 51.5144 & -28.7968 & no & no & 0.0014 & 0.4914 & 0.4865 & 0.1904 & 0.1895 & 0.1467 & 0.0013 & 0.0037 & 0.0160 & 0.1792 & 0.0663 & 0.1821 & 0.0000 & 0.0000 & 0.0000 & 0.0255 & 0.0000 & 0.0000 \\
J032603.50-284637.0 & 51.5146 & -28.7770 & no & no & 0.0074 & 0.4144 & 0.5382 & 0.1418 & 0.1515 & 0.1166 & 0.0047 & 0.0010 & 0.0337 & 0.1284 & 0.2198 & 0.0694 & 0.0720 & 0.0000 & 0.0000 & 0.6240 & 0.6681 & 0.6704 \\
J032603.60-284627.4 & 51.5150 & -28.7743 & no & no & 0.0012 & 0.4578 & 0.5165 & 0.0850 & 0.0904 & 0.0484 & 0.0008 & 0.0006 & 0.0374 & 0.1053 & 0.1424 & 0.0807 & 0.1231 & 0.0876 & 0.0000 & 0.8517 & 0.7532 & 0.7019 \\
J032603.63-283840.5 & 51.5151 & -28.6446 & no & no & 0.0021 & 0.4153 & 0.5577 & 0.1678 & 0.1746 & 0.1323 & 0.0021 & 0.0073 & 0.0386 & 0.1482 & 0.0403 & 0.1210 & 0.0000 & 0.0532 & 0.0000 & 0.0000 & 0.0302 & 0.0000 \\
J032603.66-283822.8 & 51.5153 & -28.6397 & no & no & 0.0001 & 0.4752 & 0.5009 & 0.0864 & 0.0861 & 0.0613 & 0.0001 & 0.0004 & 0.0038 & 0.0854 & 0.0447 & 0.0514 & 0.0000 & 0.0000 & 0.0000 & 0.0000 & 0.0000 & 0.0000 \\
J032603.75-284014.1 & 51.5156 & -28.6706 & no & no & 0.0547 & 0.3408 & 0.5388 & 0.4889 & 0.5242 & 0.7301 & 0.0542 & 0.2712 & 0.2318 & 0.5026 & 0.5631 & 0.5032 & 0.0595 & 0.0545 & 0.0000 & 0.4289 & 0.0789 & 0.1420 \\
      \hline
    \end{tabular}
  \end{table*}

  \begin{table*}
    \caption{Scores output by our trained classifiers for SWIRE~ELAIS-S1 candidate host galaxies. Columns are defined in \autoref{app:scores}. Full table electronic.}
    \label{tab:elais-scores}
    \begin{tabular}{ccccccccccccccccccccccc}
      \hline
SWIRE & RA & Dec & Expert host & RGZ host & CNN(Norris / All) & CNN(Norris / Compact) & CNN(Norris / Resolved) & CNN(RGZ / All) & CNN(RGZ / Compact) & CNN(RGZ / Resolved) & LR(Norris / All) & LR(Norris / Compact) & LR(Norris / Resolved) & LR(RGZ / All) & LR(RGZ / Compact) & LR(RGZ / Resolved) & RF(Norris / All) & RF(Norris / Compact) & RF(Norris / Resolved) & RF(RGZ / All) & RF(RGZ / Compact) & RF(RGZ / Resolved) \\
      \hline
J002925.73-440256.2 & 7.3572 & -44.0490 & yes & no & 0.9537 & 0.8638 & 0.5552 & 0.9195 & 0.9037 & 0.9371 & 0.9722 & 0.9955 & 0.8769 & 0.9933 & 0.9934 & 0.9658 & 0.8824 & 0.9664 & 0.7950 & 0.8078 & 0.9227 & 0.7677 \\
J002926.14-440249.0 & 7.3590 & -44.0470 & no & no & 0.7361 & 0.8752 & 0.5640 & 0.7740 & 0.7474 & 0.7952 & 0.4669 & 0.0111 & 0.4249 & 0.3926 & 0.2220 & 0.5947 & 0.2077 & 0.0000 & 0.1613 & 0.1876 & 0.0852 & 0.4546 \\
J002926.52-440247.0 & 7.3605 & -44.0464 & no & no & 0.3390 & 0.8338 & 0.5556 & 0.7275 & 0.6894 & 0.7197 & 0.2264 & 0.0254 & 0.2389 & 0.6275 & 0.3033 & 0.6812 & 0.1347 & 0.0857 & 0.0399 & 0.3582 & 0.4854 & 0.5347 \\
J002926.63-440301.1 & 7.3610 & -44.0503 & no & no & 0.2108 & 0.8251 & 0.5623 & 0.3434 & 0.3306 & 0.3292 & 0.0603 & 0.0007 & 0.0734 & 0.0688 & 0.0141 & 0.1581 & 0.0917 & 0.0000 & 0.0399 & 0.2846 & 0.1245 & 0.2833 \\
J002927.13-440232.6 & 7.3631 & -44.0424 & no & no & 0.0339 & 0.8479 & 0.5669 & 0.5853 & 0.5148 & 0.5159 & 0.0248 & 0.0334 & 0.0301 & 0.5735 & 0.5065 & 0.5265 & 0.1977 & 0.1507 & 0.0000 & 0.3334 & 0.6593 & 0.3995 \\
J002927.28-440245.3 & 7.3637 & -44.0459 & no & no & 0.0406 & 0.8345 & 0.5540 & 0.2702 & 0.2340 & 0.2133 & 0.0173 & 0.0016 & 0.0359 & 0.1056 & 0.0492 & 0.1456 & 0.0000 & 0.0000 & 0.0000 & 0.0000 & 0.0000 & 0.0287 \\
J002927.44-440238.5 & 7.3644 & -44.0440 & no & no & 0.0116 & 0.8267 & 0.5746 & 0.2228 & 0.2182 & 0.2028 & 0.0064 & 0.0049 & 0.0187 & 0.1981 & 0.1534 & 0.1493 & 0.0000 & 0.0000 & 0.0000 & 0.1565 & 0.1634 & 0.1284 \\
J002928.08-440230.3 & 7.3670 & -44.0418 & no & no & 0.0024 & 0.8626 & 0.5791 & 0.2297 & 0.1963 & 0.1549 & 0.0020 & 0.0005 & 0.0239 & 0.1337 & 0.1001 & 0.1310 & 0.0000 & 0.0000 & 0.0358 & 0.0000 & 0.0000 & 0.0190 \\
J002928.11-440312.7 & 7.3671 & -44.0535 & no & no & 0.0011 & 0.8159 & 0.5514 & 0.0377 & 0.0384 & 0.0271 & 0.0008 & 0.0013 & 0.0119 & 0.0280 & 0.0361 & 0.0205 & 0.1171 & 0.0000 & 0.0000 & 0.0873 & 0.0383 & 0.0000 \\
J002928.80-440306.8 & 7.3700 & -44.0519 & no & no & 0.0003 & 0.8405 & 0.5668 & 0.0236 & 0.0226 & 0.0136 & 0.0004 & 0.0014 & 0.0095 & 0.0339 & 0.0408 & 0.0136 & 0.0000 & 0.0000 & 0.0000 & 0.1114 & 0.1480 & 0.1584 \\
      \hline
    \end{tabular}
  \end{table*}

\section{ATLAS component cross-identifications}\label{app:xids}
  
  This section contains cross-identifications predicted by our method for each
  ATLAS radio component in CDFS and ELAIS-S1. Cross-identifications for
  ATLAS~CDFS components are shown in \autoref{tab:cdfs-xids} and
  cross-identifications for ATLAS~ELAIS-S1 are shown in
  \autoref{tab:elais-xids}. For CDFS, the cross-identification for a component
  in a quadrant is predicted using our method with binary classifiers trained
  on all other quadrants. For ELAIS-S1, we show the cross-identifications
  predicted by our method using binary classifiers trained on each CDFS
  quadrant. For CDFS, we also show the Radio Galaxy Zoo consensus, which is a
  proxy for the difficulty of cross-identifying a component (Wong et al. in
  prep).

  The columns of the cross-identification tables are defined as follows:
  \begin{itemize}
    \item \emph{ATLAS} --- ATLAS designation for radio component.
    \item \emph{RA} --- Right ascension of radio component (J2000).
    \item \emph{Dec} --- Declination of radio component (J2000).
    \item \emph{CID} --- Radio Galaxy Zoo component ID.
    \item \emph{Zooniverse ID} --- Radio Galaxy Zoo Zooniverse ID.
    \item \emph{Norris/Middelberg} --- Designation of SWIRE cross-identification from \citet{norris06} or \citet{middelberg08} for CDFS and ELAIS-S1 respectively.
    \item \emph{Norris/Middelberg RA} --- Right ascension of SWIRE cross-identification from \citet{norris06} or \citet{middelberg08} for CDFS and ELAIS-S1 respectively.
    \item \emph{Norris/Middelberg Dec} --- Right ascension of SWIRE cross-identification from \citet{norris06} or \citet{middelberg08} for CDFS and ELAIS-S1 respectively.
    \item \emph{RGZ} --- Designation of SWIRE cross-identification from Radio Galaxy Zoo (Wong et al. in prep).
    \item \emph{RGZ RA} --- Right ascension of SWIRE cross-identification from Radio Galaxy Zoo (Wong et al. in prep).
    \item \emph{RGZ Dec} --- Right ascension of SWIRE cross-identification from Radio Galaxy Zoo (Wong et al. in prep).
    \item \emph{RGZ radio consensus} --- Percentage agreement of Radio Galaxy Zoo volunteers on the radio component configuration.
    \item \emph{RGZ IR consensus} --- Percentage agreement of Radio Galaxy Zoo volunteers on the host galaxy of this radio component.
    \item \emph{$C$($L$ / $D$)} --- Designation of SWIRE cross-identification made by our method using classification model $C$ trained on label set $L$ of $D$ candidate host galaxies. $C$ may be `CNN', `LR' or `RF' for CNN, logistic regression or random forests respectively. $L$ may be `Norris' or `RGZ' for expert and Radio Galaxy Zoo labels respectively. $D$ may be `All', `Compact' or `Resolved' for each respective subset defined in \autoref{sec:experimental-setup}.
    \item \emph{$C$($L$ / $D$) RA} --- Right ascension (J2000) of SWIRE cross-identification made by our method using classification model $C$ trained on label set $L$ of $D$ candidate host galaxies. $C$, $L$ and $D$ are defined as for designation.
    \item \emph{$C$($L$ / $D$) Dec} --- Declination (J2000) of SWIRE cross-identification made by our method using classification model $C$ trained on label set $L$ of $D$ candidate host galaxies. $C$, $L$ and $D$ are defined as for designation.
  \end{itemize}

  \begin{table*}
    \caption{Cross-identifications for ATLAS~CDFS components. Columns are defined in \autoref{app:xids}. Full table electronic.}
    \label{tab:cdfs-xids}
    \begin{tabular}{ccccccccccccccccccccccccccccccccccccccccccccccccc}
      \hline
ATLAS & RA & Dec & CID & Zooniverse ID & Norris & Norris RA & Norris Dec & RGZ & RGZ RA & RGZ Dec & RGZ radio consensus & RGZ IR consensus & CNN(Norris / Compact) & CNN(Norris / Compact) RA & CNN(Norris / Compact) Dec & CNN(Norris / Resolved) & CNN(Norris / Resolved) RA & CNN(Norris / Resolved) Dec & CNN(RGZ / Compact) & CNN(RGZ / Compact) RA & CNN(RGZ / Compact) Dec & CNN(RGZ / Resolved) & CNN(RGZ / Resolved) RA & CNN(RGZ / Resolved) Dec & LR(Norris / Compact) & LR(Norris / Compact) RA & LR(Norris / Compact) Dec & LR(Norris / Resolved) & LR(Norris / Resolved) RA & LR(Norris / Resolved) Dec & LR(RGZ / Compact) & LR(RGZ / Compact) RA & LR(RGZ / Compact) Dec & LR(RGZ / Resolved) & LR(RGZ / Resolved) RA & LR(RGZ / Resolved) Dec & RF(Norris / Compact) & RF(Norris / Compact) RA & RF(Norris / Compact) Dec & RF(Norris / Resolved) & RF(Norris / Resolved) RA & RF(Norris / Resolved) Dec & RF(RGZ / Compact) & RF(RGZ / Compact) RA & RF(RGZ / Compact) Dec & RF(RGZ / Resolved) & RF(RGZ / Resolved) RA & RF(RGZ / Resolved) Dec \\
      \hline
J032602.82-284708.1C & 51.5117 & -28.7856 & CI0412 & ARG0003rb2 & J032603.15-284708.5 & 51.5132 & -28.7857 &  &  &  & 0.4516 & 0.3214 & J032602.36-284711.5 & 51.5098 & -28.7865 & J032602.36-284711.5 & 51.5098 & -28.7865 & J032602.36-284711.5 & 51.5098 & -28.7865 & J032602.36-284711.5 & 51.5098 & -28.7865 & J032604.58-284650.9 & 51.5191 & -28.7808 & J032602.08-284713.1 & 51.5087 & -28.787 & J032602.36-284711.5 & 51.5098 & -28.7865 & J032602.36-284711.5 & 51.5098 & -28.7865 & J032603.15-284708.5 & 51.5132 & -28.7857 & J032602.36-284711.5 & 51.5098 & -28.7865 & J032602.36-284711.5 & 51.5098 & -28.7865 & J032602.36-284711.5 & 51.5098 & -28.7865 \\
J032615.49-284629.4C & 51.5646 & -28.7749 & CI0614 & ARG0003rfr & J032615.41-284630.7 & 51.5642 & -28.7752 & J032615.41-284630.7 & 51.5642 & -28.7752 & 0.2941 & 0.8000 & J032615.41-284630.7 & 51.5642 & -28.7752 & J032615.41-284630.7 & 51.5642 & -28.7752 & J032615.41-284630.7 & 51.5642 & -28.7752 & J032615.41-284630.7 & 51.5642 & -28.7752 & J032615.41-284630.7 & 51.5642 & -28.7752 & J032615.41-284630.7 & 51.5642 & -28.7752 & J032615.41-284630.7 & 51.5642 & -28.7752 & J032615.41-284630.7 & 51.5642 & -28.7752 & J032615.41-284630.7 & 51.5642 & -28.7752 & J032615.41-284630.7 & 51.5642 & -28.7752 & J032615.41-284630.7 & 51.5642 & -28.7752 & J032615.41-284630.7 & 51.5642 & -28.7752 \\
J032615.55-280559.8C & 51.5648 & -28.1000 & CI0320 & ARG0003r8s & J032615.52-280559.8 & 51.5647 & -28.1000 & J032615.52-280559.8 & 51.5647 & -28.1000 & 0.5625 & 0.8333 & J032615.52-280559.8 & 51.5647 & -28.1000 & J032615.52-280559.8 & 51.5647 & -28.1000 & J032615.52-280559.8 & 51.5647 & -28.1000 & J032615.52-280559.8 & 51.5647 & -28.1000 & J032615.52-280559.8 & 51.5647 & -28.1000 & J032615.52-280559.8 & 51.5647 & -28.1000 & J032615.52-280559.8 & 51.5647 & -28.1000 & J032615.52-280559.8 & 51.5647 & -28.1000 & J032615.52-280559.8 & 51.5647 & -28.1000 & J032615.52-280559.8 & 51.5647 & -28.1000 & J032615.52-280559.8 & 51.5647 & -28.1000 & J032615.52-280559.8 & 51.5647 & -28.1000 \\
J032617.35-280710.2C & 51.5723 & -28.1195 & CI0059C1 & ARG0003r2j & J032617.89-280707.2 & 51.5746 & -28.1187 & J032617.89-280707.2 & 51.5746 & -28.1187 & 0.4146 & 1.0000 & J032617.89-280707.2 & 51.5746 & -28.1187 & J032617.89-280707.2 & 51.5746 & -28.1187 & J032617.89-280707.2 & 51.5746 & -28.1187 & J032617.89-280707.2 & 51.5746 & -28.1187 & J032615.86-280628.8 & 51.5661 & -28.1080 & J032615.16-280742.2 & 51.5632 & -28.1284 & J032615.86-280628.8 & 51.5661 & -28.1080 & J032618.84-280722.6 & 51.5785 & -28.1230 & J032617.89-280707.2 & 51.5746 & -28.1187 & J032617.89-280707.2 & 51.5746 & -28.1187 & J032617.89-280707.2 & 51.5746 & -28.1187 & J032617.89-280707.2 & 51.5746 & -28.1187 \\
J032625.13-280909.8C & 51.6047 & -28.1527 & CI0409 & ARG0003raz & J032625.19-280910.1 & 51.6050 & -28.1528 & J032625.19-280910.1 & 51.6050 & -28.1528 & 0.3158 & 0.6667 & J032625.19-280910.1 & 51.6050 & -28.1528 & J032625.19-280910.1 & 51.6050 & -28.1528 & J032624.50-280905.9 & 51.6021 & -28.1517 & J032625.19-280910.1 & 51.6050 & -28.1528 & J032625.19-280910.1 & 51.6050 & -28.1528 & J032625.19-280910.1 & 51.6050 & -28.1528 & J032625.19-280910.1 & 51.6050 & -28.1528 & J032625.19-280910.1 & 51.6050 & -28.1528 & J032625.19-280910.1 & 51.6050 & -28.1528 & J032625.19-280910.1 & 51.6050 & -28.1528 & J032625.19-280910.1 & 51.6050 & -28.1528 & J032625.19-280910.1 & 51.6050 & -28.1528 \\
J032629.10-280650.1C & 51.6213 & -28.1139 & CI0963 & ARG0003ro4 & J032629.13-280650.7 & 51.6214 & -28.1141 & J032626.74-280636.7 & 51.6114 & -28.1102 & 0.3333 & 1.0000 & J032629.13-280650.7 & 51.6214 & -28.1141 & J032629.13-280650.7 & 51.6214 & -28.1141 & J032629.13-280650.7 & 51.6214 & -28.1141 & J032629.13-280650.7 & 51.6214 & -28.1141 & J032629.13-280650.7 & 51.6214 & -28.1141 & J032629.13-280650.7 & 51.6214 & -28.1141 & J032629.13-280650.7 & 51.6214 & -28.1141 & J032629.13-280650.7 & 51.6214 & -28.1141 & J032629.13-280650.7 & 51.6214 & -28.1141 & J032629.13-280650.7 & 51.6214 & -28.1141 & J032629.13-280650.7 & 51.6214 & -28.1141 & J032629.13-280650.7 & 51.6214 & -28.1141 \\
J032629.61-284052.7C & 51.6234 & -28.6813 & CI0304 & ARG0003r8e & J032629.54-284055.8 & 51.6231 & -28.6822 & J032629.54-284055.8 & 51.6231 & -28.6822 & 0.2676 & 1.0000 & J032629.54-284051.9 & 51.6231 & -28.6811 & J032629.54-284051.9 & 51.6231 & -28.6811 & J032629.54-284051.9 & 51.6231 & -28.6811 & J032629.54-284051.9 & 51.6231 & -28.6811 & J032629.54-284051.9 & 51.6231 & -28.6811 & J032629.54-284051.9 & 51.6231 & -28.6811 & J032629.54-284051.9 & 51.6231 & -28.6811 & J032629.54-284051.9 & 51.6231 & -28.6811 & J032629.54-284051.9 & 51.6231 & -28.6811 & J032629.54-284051.9 & 51.6231 & -28.6811 & J032629.54-284051.9 & 51.6231 & -28.6811 & J032629.54-284051.9 & 51.6231 & -28.6811 \\
J032629.92-284753.5C & 51.6247 & -28.7982 & CI0120 & ARG0003r3w & J032629.81-284754.4 & 51.6242 & -28.7985 & J032629.81-284754.4 & 51.6242 & -28.7985 & 1.0000 & 0.8571 & J032629.81-284754.4 & 51.6242 & -28.7985 & J032629.81-284754.4 & 51.6242 & -28.7985 & J032629.81-284754.4 & 51.6242 & -28.7985 & J032629.81-284754.4 & 51.6242 & -28.7985 & J032629.81-284754.4 & 51.6242 & -28.7985 & J032629.81-284754.4 & 51.6242 & -28.7985 & J032629.81-284754.4 & 51.6242 & -28.7985 & J032629.81-284754.4 & 51.6242 & -28.7985 & J032630.12-284751.2 & 51.6255 & -28.7976 & J032629.81-284754.4 & 51.6242 & -28.7985 & J032629.81-284754.4 & 51.6242 & -28.7985 & J032629.81-284754.4 & 51.6242 & -28.7985 \\
J032630.66-283657.3C & 51.6278 & -28.6159 & CI0172C1 & ARG0003r55 & J032630.64-283658.0 & 51.6277 & -28.6161 & J032628.56-283744.8 & 51.619 & -28.6291 & 0.3611 & 0.7308 & J032630.64-283658.0 & 51.6277 & -28.6161 & J032630.64-283658.0 & 51.6277 & -28.6161 & J032630.64-283658.0 & 51.6277 & -28.6161 & J032630.64-283658.0 & 51.6277 & -28.6161 & J032630.64-283658.0 & 51.6277 & -28.6161 & J032630.64-283658.0 & 51.6277 & -28.6161 & J032630.64-283658.0 & 51.6277 & -28.6161 & J032630.64-283658.0 & 51.6277 & -28.6161 & J032630.64-283658.0 & 51.6277 & -28.6161 & J032630.64-283658.0 & 51.6277 & -28.6161 & J032630.64-283658.0 & 51.6277 & -28.6161 & J032630.64-283658.0 & 51.6277 & -28.6161 \\
J032634.59-282022.8C & 51.6441 & -28.3397 & CI0757 & ARG0003rj2 & J032634.58-282022.8 & 51.6441 & -28.3397 & J032631.96-281941.0 & 51.6332 & -28.3281 & 0.5781 & 0.5405 & J032634.58-282022.8 & 51.6441 & -28.3397 & J032634.58-282022.8 & 51.6441 & -28.3397 & J032634.58-282022.8 & 51.6441 & -28.3397 & J032634.58-282022.8 & 51.6441 & -28.3397 & J032634.58-282022.8 & 51.6441 & -28.3397 & J032634.58-282022.8 & 51.6441 & -28.3397 & J032634.58-282022.8 & 51.6441 & -28.3397 & J032634.58-282022.8 & 51.6441 & -28.3397 & J032634.58-282022.8 & 51.6441 & -28.3397 & J032634.58-282022.8 & 51.6441 & -28.3397 & J032634.58-282022.8 & 51.6441 & -28.3397 & J032634.58-282022.8 & 51.6441 & -28.3397 \\
      \hline
    \end{tabular}
  \end{table*}

  \begin{table*}
    \caption{Cross-identifications for ATLAS~ELAIS-S1 components. Columns are defined in \autoref{app:xids}. Full table electronic.}
    \label{tab:elais-xids}
    \begin{tabular}{ccccccccccccccccccccccccccccccccccccccccccccccccccccccccccccccccccc}
      \hline
ATLAS & RA & Dec & CID & Zooniverse ID & Middelberg & Middelberg RA & Middelberg Dec & RGZ & RGZ RA & RGZ Dec & RGZ radio consensus & RGZ IR consensus & CNN(Norris / All) & CNN(Norris / All) RA & CNN(Norris / All) Dec & CNN(Norris / Compact) & CNN(Norris / Compact) RA & CNN(Norris / Compact) Dec & CNN(Norris / Resolved) & CNN(Norris / Resolved) RA & CNN(Norris / Resolved) Dec & CNN(RGZ / All) & CNN(RGZ / All) RA & CNN(RGZ / All) Dec & CNN(RGZ / Compact) & CNN(RGZ / Compact) RA & CNN(RGZ / Compact) Dec & CNN(RGZ / Resolved) & CNN(RGZ / Resolved) RA & CNN(RGZ / Resolved) Dec & LR(Norris / All) & LR(Norris / All) RA & LR(Norris / All) Dec & LR(Norris / Compact) & LR(Norris / Compact) RA & LR(Norris / Compact) Dec & LR(Norris / Resolved) & LR(Norris / Resolved) RA & LR(Norris / Resolved) Dec & LR(RGZ / All) & LR(RGZ / All) RA & LR(RGZ / All) Dec & LR(RGZ / Compact) & LR(RGZ / Compact) RA & LR(RGZ / Compact) Dec & LR(RGZ / Resolved) & LR(RGZ / Resolved) RA & LR(RGZ / Resolved) Dec & RF(Norris / All) & RF(Norris / All) RA & RF(Norris / All) Dec & RF(Norris / Compact) & RF(Norris / Compact) RA & RF(Norris / Compact) Dec & RF(Norris / Resolved) & RF(Norris / Resolved) RA & RF(Norris / Resolved) Dec & RF(RGZ / All) & RF(RGZ / All) RA & RF(RGZ / All) Dec & RF(RGZ / Compact) & RF(RGZ / Compact) RA & RF(RGZ / Compact) Dec & RF(RGZ / Resolved) & RF(RGZ / Resolved) RA & RF(RGZ / Resolved) Dec \\
      \hline
J002925.68-440256.8 & 7.3570 & -44.0491 & C0375 &  & J002925.73-440256.2 & 7.3572 & -44.0490 &  &  &  &  &  & J002925.73-440256.2 & 7.3572 & -44.049 & J002925.73-440256.2 & 7.3572 & -44.0490 & J002925.73-440256.2 & 7.3572 & -44.0490 & J002925.73-440256.2 & 7.3572 & -44.0490 & J002925.73-440256.2 & 7.3572 & -44.0490 & J002925.73-440256.2 & 7.3572 & -44.0490 & J002925.73-440256.2 & 7.3572 & -44.0490 & J002925.73-440256.2 & 7.3572 & -44.0490 & J002925.73-440256.2 & 7.3572 & -44.0490 & J002925.73-440256.2 & 7.3572 & -44.0490 & J002925.73-440256.2 & 7.3572 & -44.0490 & J002925.73-440256.2 & 7.3572 & -44.0490 & J002925.73-440256.2 & 7.3572 & -44.0490 & J002925.73-440256.2 & 7.3572 & -44.0490 & J002925.73-440256.2 & 7.3572 & -44.0490 & J002925.73-440256.2 & 7.3572 & -44.0490 & J002925.73-440256.2 & 7.3572 & -44.0490 & J002925.73-440256.2 & 7.3572 & -44.0490 \\
J002938.19-432946.7 & 7.4092 & -43.4963 & C0832 &  & J002938.07-432947.9 & 7.4087 & -43.4967 &  &  &  &  &  & J002938.07-432947.9 & 7.4087 & -43.4967 & J002938.07-432947.9 & 7.4087 & -43.4967 & J002938.07-432947.9 & 7.4087 & -43.4967 & J002937.50-432945.4 & 7.4063 & -43.4959 & J002937.50-432945.4 & 7.4063 & -43.4959 & J002937.50-432945.4 & 7.4063 & -43.4959 & J002938.07-432947.9 & 7.4087 & -43.4967 & J002938.07-432947.9 & 7.4087 & -43.4967 & J002938.07-432947.9 & 7.4087 & -43.4967 & J002938.07-432947.9 & 7.4087 & -43.4967 & J002938.07-432947.9 & 7.4087 & -43.4967 & J002938.07-432947.9 & 7.4087 & -43.4967 & J002938.07-432947.9 & 7.4087 & -43.4967 & J002938.07-432947.9 & 7.4087 & -43.4967 & J002938.07-432947.9 & 7.4087 & -43.4967 & J002938.07-432947.9 & 7.4087 & -43.4967 & J002938.07-432947.9 & 7.4087 & -43.4967 & J002938.07-432947.9 & 7.4087 & -43.4967 \\
J002940.13-440309.2 & 7.4172 & -44.0526 & C0374 &  & J002940.19-440309.6 & 7.4175 & -44.0527 &  &  &  &  &  & J002940.19-440309.6 & 7.4175 & -44.0527 & J002940.19-440309.6 & 7.4175 & -44.0527 & J002940.19-440309.6 & 7.4175 & -44.0527 & J002940.19-440309.6 & 7.4175 & -44.0527 & J002940.19-440309.6 & 7.4175 & -44.0527 & J002940.19-440309.6 & 7.4175 & -44.0527 & J002940.19-440309.6 & 7.4175 & -44.0527 & J002940.19-440309.6 & 7.4175 & -44.0527 & J002940.19-440309.6 & 7.4175 & -44.0527 & J002940.19-440309.6 & 7.4175 & -44.0527 & J002940.19-440309.6 & 7.4175 & -44.0527 & J002940.19-440309.6 & 7.4175 & -44.0527 & J002940.19-440309.6 & 7.4175 & -44.0527 & J002940.19-440309.6 & 7.4175 & -44.0527 & J002940.19-440309.6 & 7.4175 & -44.0527 & J002940.19-440309.6 & 7.4175 & -44.0527 & J002940.19-440309.6 & 7.4175 & -44.0527 & J002940.19-440309.6 & 7.4175 & -44.0527 \\
J002943.14-440812.3 & 7.4298 & -44.1368 & C0302 &  & J002943.15-440813.6 & 7.4298 & -44.1371 &  &  &  &  &  & J002943.15-440813.6 & 7.4298 & -44.1371 & J002943.15-440813.6 & 7.4298 & -44.1371 & J002943.15-440813.6 & 7.4298 & -44.1371 & J002943.15-440813.6 & 7.4298 & -44.1371 & J002943.15-440813.6 & 7.4298 & -44.1371 & J002943.15-440813.6 & 7.4298 & -44.1371 & J002943.15-440813.6 & 7.4298 & -44.1371 & J002943.15-440813.6 & 7.4298 & -44.1371 & J002943.15-440813.6 & 7.4298 & -44.1371 & J002943.15-440813.6 & 7.4298 & -44.1371 & J002943.15-440813.6 & 7.4298 & -44.1371 & J002943.15-440813.6 & 7.4298 & -44.1371 & J002943.15-440813.6 & 7.4298 & -44.1371 & J002943.15-440813.6 & 7.4298 & -44.1371 & J002943.15-440813.6 & 7.4298 & -44.1371 & J002943.15-440813.6 & 7.4298 & -44.1371 & J002943.15-440813.6 & 7.4298 & -44.1371 & J002943.15-440813.6 & 7.4298 & -44.1371 \\
J002944.51-433627.8 & 7.4355 & -43.6077 & C0727 &  & J002944.36-433630.2 & 7.4348 & -43.6084 &  &  &  &  &  & J002944.36-433630.2 & 7.4348 & -43.6084 & J002944.36-433630.2 & 7.4348 & -43.6084 & J002944.36-433630.2 & 7.43484 & -43.6084 & J002944.36-433630.2 & 7.4348 & -43.6084 & J002944.36-433630.2 & 7.43484 & -43.6084 & J002944.36-433630.2 & 7.4348 & -43.6084 & J002944.36-433630.2 & 7.43484 & -43.6084 & J002944.36-433630.2 & 7.4348 & -43.6084 & J002944.36-433630.2 & 7.43484 & -43.6084 & J002944.36-433630.2 & 7.4348 & -43.6084 & J002944.36-433630.2 & 7.43484 & -43.6084 & J002944.36-433630.2 & 7.4348 & -43.6084 & J002944.36-433630.2 & 7.43484 & -43.6084 & J002944.36-433630.2 & 7.4348 & -43.6084 & J002944.36-433630.2 & 7.43484 & -43.6084 & J002944.36-433630.2 & 7.4348 & -43.6084 & J002944.36-433630.2 & 7.43484 & -43.6084 & J002944.36-433630.2 & 7.4348 & -43.6084 \\
J002945.31-432148.5 & 7.4388 & -43.3635 & C0943.1 &  & J002945.64-432149.3 & 7.4402 & -43.3637 &  &  &  &  &  & J002945.64-432149.3 & 7.4402 & -43.3637 & J002945.64-432149.3 & 7.4402 & -43.3637 & J002945.64-432149.3 & 7.4402 & -43.3637 & J002945.64-432149.3 & 7.4402 & -43.3637 & J002945.64-432149.3 & 7.4402 & -43.3637 & J002945.64-432149.3 & 7.4402 & -43.3637 & J002945.64-432149.3 & 7.4402 & -43.3637 & J002945.64-432149.3 & 7.4402 & -43.3637 & J002945.64-432149.3 & 7.4402 & -43.3637 & J002945.64-432149.3 & 7.4402 & -43.3637 & J002945.64-432149.3 & 7.4402 & -43.3637 & J002945.64-432149.3 & 7.4402 & -43.3637 & J002945.64-432149.3 & 7.4402 & -43.3637 & J002945.64-432149.3 & 7.4402 & -43.3637 & J002945.64-432149.3 & 7.4402 & -43.3637 & J002945.64-432149.3 & 7.4402 & -43.3637 & J002945.64-432149.3 & 7.4402 & -43.3637 & J002945.64-432149.3 & 7.4402 & -43.3637 \\
J002946.14-432149.1 & 7.4423 & -43.3637 & C0943 &  & J002945.64-432149.3 & 7.4402 & -43.3637 &  &  &  &  &  & J002945.64-432149.3 & 7.4402 & -43.3637 & J002945.64-432149.3 & 7.4402 & -43.3637 & J002945.64-432149.3 & 7.4402 & -43.3637 & J002945.64-432149.3 & 7.4402 & -43.3637 & J002945.64-432149.3 & 7.4402 & -43.3637 & J002945.64-432149.3 & 7.4402 & -43.3637 & J002945.64-432149.3 & 7.4402 & -43.3637 & J002945.64-432149.3 & 7.4402 & -43.3637 & J002945.64-432149.3 & 7.4402 & -43.3637 & J002945.64-432149.3 & 7.4402 & -43.3637 & J002945.64-432149.3 & 7.4402 & -43.3637 & J002945.64-432149.3 & 7.4402 & -43.3637 & J002945.64-432149.3 & 7.4402 & -43.3637 & J002945.64-432149.3 & 7.4402 & -43.3637 & J002945.64-432149.3 & 7.4402 & -43.3637 & J002945.64-432149.3 & 7.4402 & -43.3637 & J002945.64-432149.3 & 7.4402 & -43.3637 & J002945.64-432149.3 & 7.4402 & -43.3637 \\
J002949.89-440541.4 & 7.4579 & -44.0948 & C0345 &  &  &  &  &  &  &  &  &  & J002951.26-440556.4 & 7.4636 & -44.099 & J002951.44-440546.1 & 7.4644 & -44.0962 & J002951.44-440546.1 & 7.4644 & -44.0962 & J002951.44-440546.1 & 7.4644 & -44.0962 & J002951.44-440546.1 & 7.4644 & -44.0962 & J002951.44-440546.1 & 7.4644 & -44.0962 & J002951.26-440556.4 & 7.4636 & -44.0990 & J002951.26-440556.4 & 7.4636 & -44.0990 & J002951.26-440556.4 & 7.4636 & -44.0990 & J002951.26-440556.4 & 7.4636 & -44.0990 & J002951.26-440556.4 & 7.4636 & -44.0990 & J002951.26-440556.4 & 7.4636 & -44.0990 & J002951.26-440556.4 & 7.4636 & -44.0990 & J002951.26-440556.4 & 7.4636 & -44.0990 & J002951.26-440556.4 & 7.4636 & -44.0990 & J002949.13-440536.5 & 7.4547 & -44.0935 & J002949.13-440536.5 & 7.4547 & -44.0935 & J002949.13-440536.5 & 7.4547 & -44.0935 \\
J002951.13-432354.3 & 7.4631 & -43.3984 & C0924 &  & J002951.14-432355.3 & 7.4631 & -43.3987 &  &  &  &  &  & J002951.14-432355.3 & 7.4631 & -43.3987 & J002951.14-432355.3 & 7.4631 & -43.3987 & J002951.14-432355.3 & 7.4631 & -43.3987 & J002951.14-432355.3 & 7.4631 & -43.3987 & J002951.14-432355.3 & 7.4631 & -43.3987 & J002951.14-432355.3 & 7.4631 & -43.3987 & J002951.14-432355.3 & 7.4631 & -43.3987 & J002951.14-432355.3 & 7.4631 & -43.3987 & J002951.14-432355.3 & 7.4631 & -43.3987 & J002951.14-432355.3 & 7.4631 & -43.3987 & J002951.14-432355.3 & 7.4631 & -43.3987 & J002951.14-432355.3 & 7.4631 & -43.3987 & J002951.14-432355.3 & 7.4631 & -43.3987 & J002951.14-432355.3 & 7.4631 & -43.3987 & J002951.14-432355.3 & 7.4631 & -43.3987 & J002951.14-432355.3 & 7.4631 & -43.3987 & J002951.14-432355.3 & 7.4631 & -43.3987 & J002951.14-432355.3 & 7.4631 & -43.3987 \\
J002951.19-440556.6 & 7.4633 & -44.0991 & C0342 &  & J002951.26-440556.4 & 7.4636 & -44.0990 &  &  &  &  &  & J002951.26-440556.4 & 7.4636 & -44.0990 & J002951.26-440556.4 & 7.4636 & -44.0990 & J002951.44-440546.1 & 7.4644 & -44.0962 & J002951.51-440617.1 & 7.4646 & -44.1048 & J002951.51-440617.1 & 7.4646 & -44.1048 & J002951.44-440546.1 & 7.4644 & -44.0962 & J002951.26-440556.4 & 7.4636 & -44.0990 & J002951.26-440556.4 & 7.4636 & -44.0990 & J002951.26-440556.4 & 7.4636 & -44.0990 & J002951.26-440556.4 & 7.4636 & -44.0990 & J002951.26-440556.4 & 7.4636 & -44.0990 & J002951.26-440556.4 & 7.4636 & -44.0990 & J002951.26-440556.4 & 7.4636 & -44.0990 & J002951.26-440556.4 & 7.4636 & -44.0990 & J002951.26-440556.4 & 7.4636 & -44.0990 & J002951.26-440556.4 & 7.4636 & -44.0990 & J002951.26-440556.4 & 7.4636 & -44.0990 & J002951.26-440556.4 & 7.4636 & -44.0990 \\
      \hline
    \end{tabular}
  \end{table*}

\section{Cross-identification figures}\label{app:examples}

    This section contains figures of cross-identifications of each ATLAS radio component in CDFS and ELAIS-S1.

    \begin{figure*}
      \centering
      \subfloat[]{\includegraphics{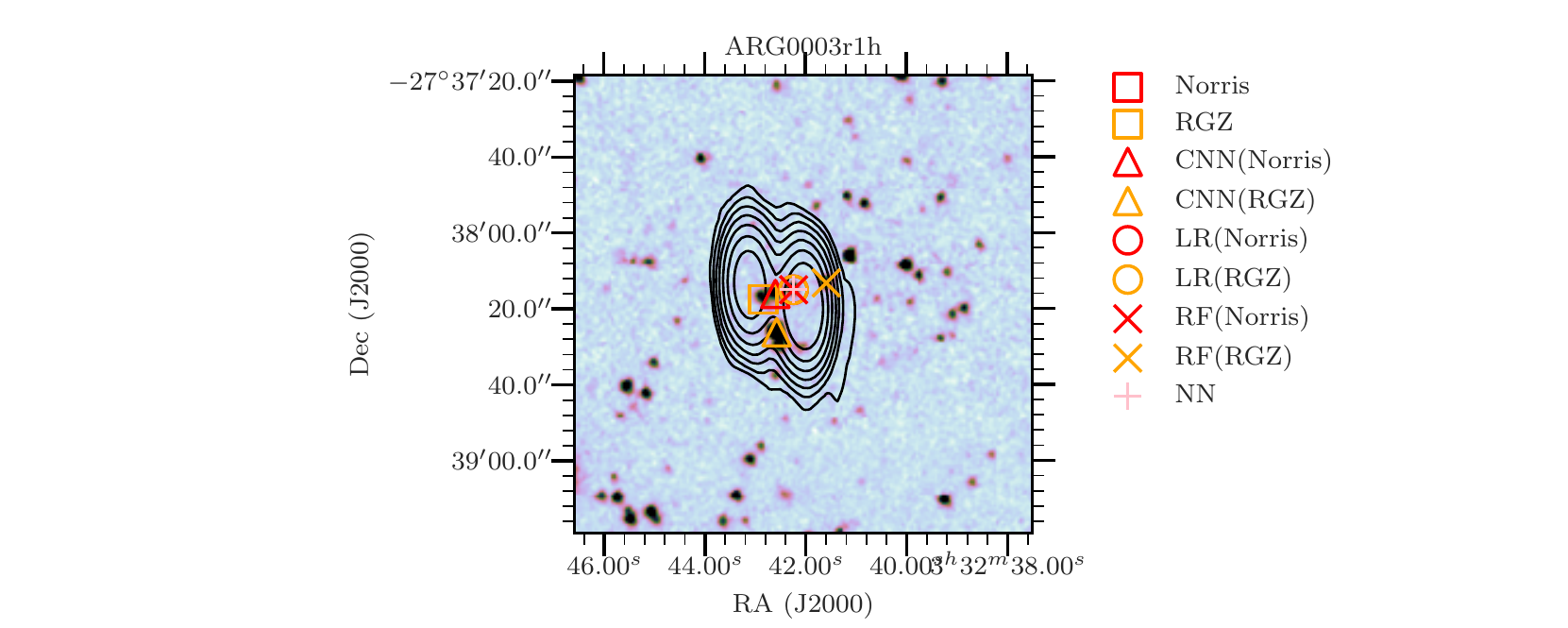}}

      \subfloat[]{\includegraphics{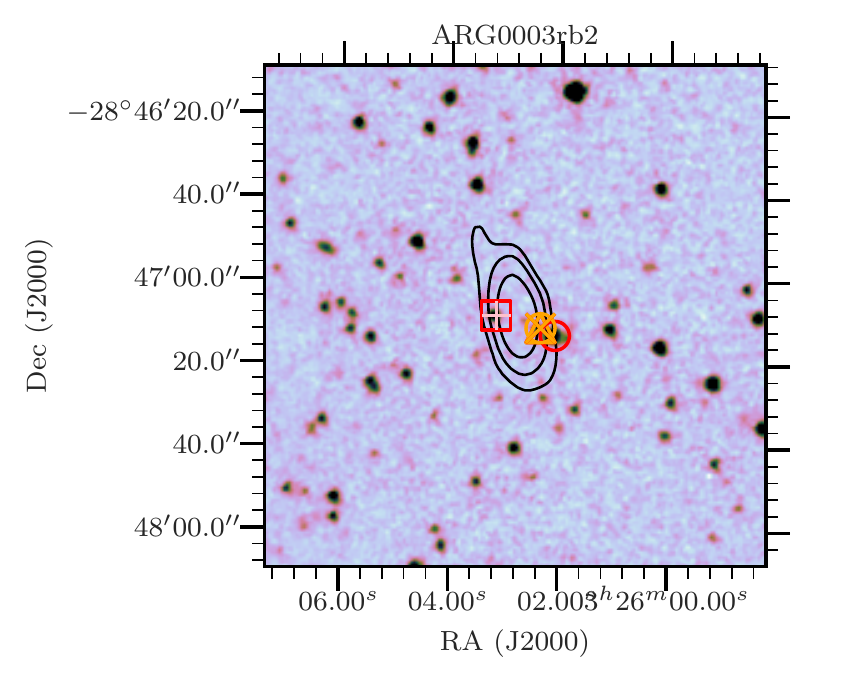}}%
      \subfloat[]{\includegraphics{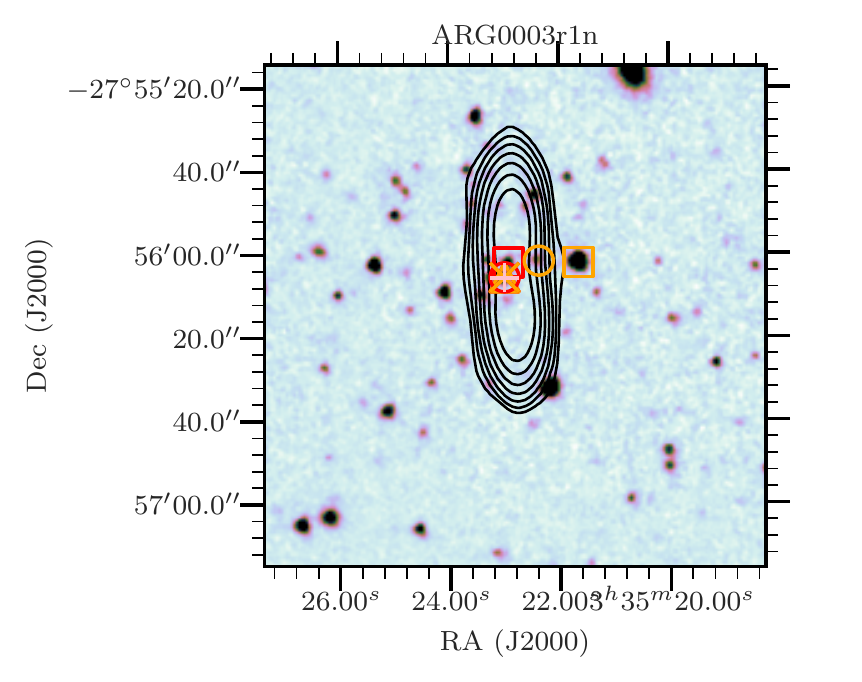}}

      \subfloat[]{\includegraphics{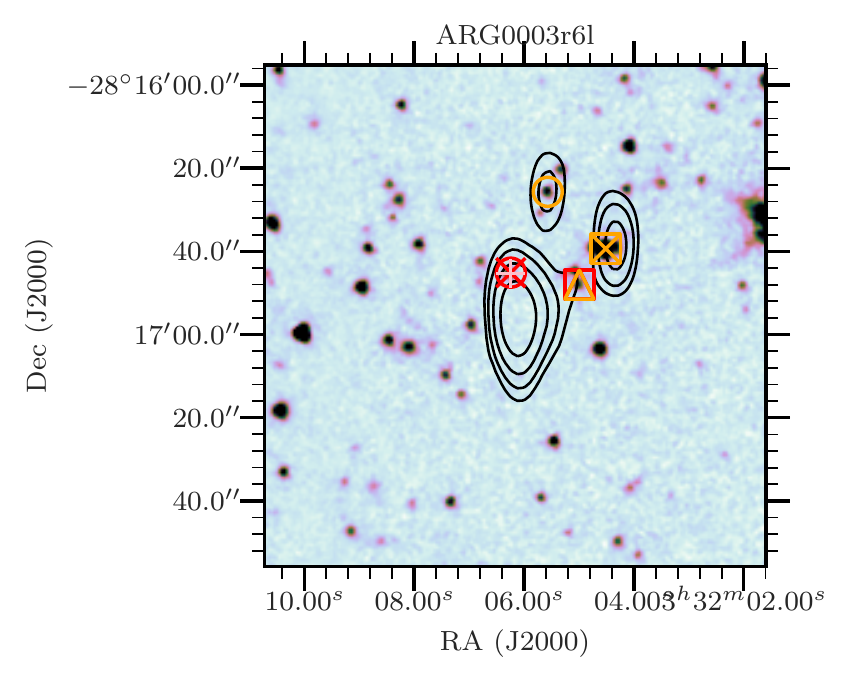}}%
      \subfloat[]{\includegraphics{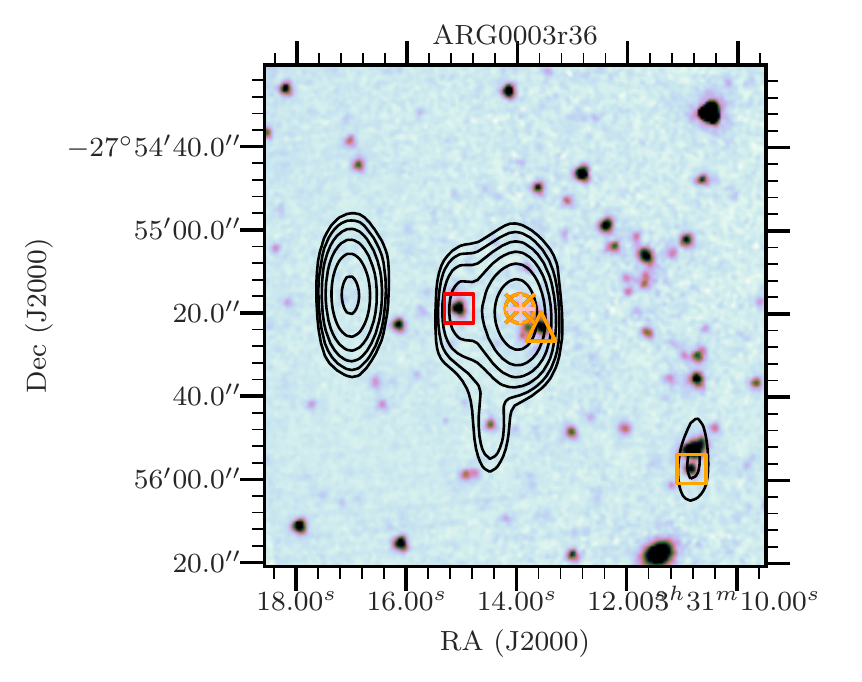}}
      \caption{\label{fig:examples} Examples of resolved sources with high disagreement between cross-identifiers. The contours show ATLAS radio data and start at $4\sigma$, increasing geometrically by a factor of 2. The background image is the \unit{3.6}{\micro\meter} SWIRE image. Binary classifier model/training set combinations are denoted $C(S)$ where $C$ is the binary classifier model and $S$ is the training set. `LR' is logistic regression, `CNN' is convolutional neural networks, and `RF' is random forests. `Norris' refers to the expert labels and `RGZ' refers to the Radio Galaxy Zoo labels. The cross-identification made by nearest-neighbours is shown by `NN'. The complete set of figures for 469 examples is available in the supplementary information.}
    \end{figure*}

\bsp	%
\label{lastpage}
\end{document}